\documentclass[aps,preprint,floatfix]{revtex4}%
\usepackage{amsfonts}
\usepackage{amsmath}
\usepackage{epsfig}
\usepackage{amssymb}
\usepackage[figuresright]{rotating}
\newcommand{\kval}{\kappa_{\rm val}}
\newcommand{\ksea}{\kappa_{\rm sea}}

\newcommand{\msea}{m^q_{\rm sea}}

\newcommand{\be}{\begin{equation}}
\newcommand{\ee}{\end{equation}}
\newcommand{\ba}{\begin{array}}
\newcommand{\ea}{\end{array}}
\newcommand{\bea}{\begin{eqnarray}}
\newcommand{\eea}{\end{eqnarray}}

\newcommand{\mev}{\,{\rm MeV}}

\newcommand{\mpi}{m_\pi}

\newcommand{\non}{\nonumber}

\newcommand{\err}[2]{${\scriptstyle {}^{+{#1}}_{-{#2}}}$}
\newcommand{\er}[2]{{\scriptstyle {}^{+{#1}}_{-{#2}}}}

\begin{document}
\vspace*{3mm}
\title[Chiral Extrapolation of pQQCD]{Unified chiral analysis of the vector meson 
spectrum from lattice QCD}

\author{W.~Armour}
\author{C.~R.~Allton}
\affiliation{Department of Physics, University of Wales Swansea, Swansea
SA2~8PP, Wales}
\author{D.~B.~Leinweber}
\affiliation{Special Research Center for the
             Subatomic Structure of Matter,
             and Department of Physics,
             University of Adelaide, Adelaide SA 5005, Australia}
\author{A.~W.~Thomas}
\author{R.~D.~Young}
\affiliation{Jefferson Lab, 12000 Jefferson Ave., Newport News VA 23606, USA
}
\begin{abstract}
The chiral extrapolation of the vector meson mass calculated in
partially-quenched lattice simulations is investigated. The leading
one-loop corrections to the vector meson mass are derived for
partially-quenched QCD. A large sample of lattice results from the
CP-PACS Collaboration is analysed, with explicit corrections for
finite lattice spacing artifacts.  To incorporate the effect of the
opening decay channel as the chiral limit is approached, the
extrapolation is studied using a necessary phenomenological
extension of chiral effective field theory.  This chiral analysis
also provides a quantitative estimate of the leading finite volume
corrections. It is found that the discretisation, finite-volume and
partial quenching effects can all be very well described in this
framework, producing an extrapolated value of $M_\rho$ in excellent
agreement with experiment. This procedure is also compared with
extrapolations based on polynomial forms, where the results are much
less enlightening.
\end{abstract}

\preprint{SWAT/05/404}
\preprint{ADP-05-20/T630}
\preprint{JLAB-THY-05-432}
\startpage{1}
\endpage{2}
\maketitle



\section{Introduction}
There has been great progress in lattice QCD in recent years,
associated both with Moore's Law and with improved algorithms, which
mean that one can work with larger lattice spacings and still
approximate the continuum limit well. The CP-PACS group has devoted
considerable effort to the study of the masses of the lowest mass
baryons and vector mesons.  This has led, for example, to a comprehensive set
of data for the mass of the $\rho$-meson in partially quenched QCD,
with exceptionally small statistical errors \cite{cppacs}.  We shall
exploit this data.

The remaining barrier to direct comparison with experimental data is
the fact that calculations take much longer as the quark mass
approaches the chiral limit. Indeed the time for a given calculation
scales somewhere in the range $m_\pi^{-7}$ to $m_\pi^{-9}$, depending
on how hard one works to preserve chiral symmetry. As a result there
has been considerable interest in using chiral perturbation theory
($\chi$PT), an effective field theory (EFT) built on the symmetries of
QCD, to provide a functional form for hadron properties as a function
of quark mass~\cite{Leinweber:2003dg,Procura:2003ig}.  In principle,
such a functional form can then be used to extrapolate from the large
pion masses where lattice data exists to the physical
value. Unfortunately, there is considerable evidence that the
convergence of dimensionally regularised $\chi$PT is too slow for this
expansion to be reliable at present
\cite{Leinweber:1999ig,Bernard:2002yk,Young:2002ib,Durr:2002zx,%
Beane:2004ks,Thomas:2004iw,Leinweber:2005xz}.

On the other hand, it can be shown that a reformulation of $\chi$PT
using finite-range regularisation (FRR) effectively resums the chiral
expansion, leaving a residual series with much better convergence
properties \cite{Leinweber:2003dg,Young:2002ib}.  The FRR expansion is
mathematically equivalent to dimensionally regularised $\chi$PT to the
finite order one is working \cite{Donoghue:1998bs,Young:2002ib}.
Systematic errors associated with the functional form of the regulator
are at the fraction of a percent level \cite{Leinweber:2003dg}.
A formal description of the formulation of baryon $\chi$PT using a
momentum cutoff (or FRR) has recently been considered by Djukanovic
{\it et al.}~\cite{Djukanovic:2005jy}.
The price of such an approach is a residual dependence on the
regulator mass, which governs the manner in which the loop integrals
vanish as the pion mass grows large.  However, if it can be
demonstrated that reasonable variation of this mass does not
significantly change the extrapolated values of physical properties,
one has made progress.  This seems to be the case for the nucleon mass
\cite{Thomas:2004iw} and magnetic moments \cite{Young:2004tb}, for
example, where ``reasonable variation'' is taken to be $\pm$20\%
around the best fit value of the optimal regulator mass.

In order to test whether the problem is indeed solved in this way one
needs a large body of accurate data. This is in fact available for the
$\rho$ meson, where CP-PACS has carried out lattice simulations of
partially quenched QCD (pQQCD) with a wide range of sea and valence
masses. This sector requires a modified effective field theory, namely
partially-quenched chiral perturbation theory (PQ$\chi$PT)
\cite{Golterman:1997st,Sharpe:2001fh}. Formal developments in this
field have made significant progress in the study of a range of
hadronic observables --- see
Refs.~\cite{Chen:2001yi,Beane:2002vq,Leinweber:2002qb,Arndt:2003ww,%
Arndt:2004bg,Bijnens:2004hk,Detmold:2005pt} for example.

In considering the $\rho$ meson, analysis of modern lattice results
requires one to extend beyond the low-energy EFT.  Near the chiral
limit, the $\rho$ decays to two energetic pions, whereas at the
quark-masses simulated on the lattice the $\rho$ is stable. The pions
contributing to the imaginary part of the $\rho$ mass cannot be
considered soft, and therefore cannot be systematically incorporated
into a low-energy counting scheme
\cite{Bijnens:1997ni,Bruns:2004tj}. Because almost all the lattice
simulation points in this analysis lie in the region $\mpi>m_\rho/2$,
it is evident that the extrapolation to the chiral regime will
encounter a threshold effect where the decay channel opens. To
incorporate this physical threshold, we model the $\rho\to\pi\pi$
self-energy diagram constrained to reproduce the observed width at the
physical pion mass. Including this contribution also provides a model
of the finite volume corrections arising from the infrared component
of the loop integral. In particular, we can also describe the lattice
results in the region $\mpi<m_\rho/2$, where the decay channel
is still energetically forbidden because of momentum discretisation.

This large body of pQQCD simulation data is analysed within a
framework which incorporates the leading low-energy behaviour of
partially-quenched EFT, together with a model for describing the decay
channel of the $\rho$ meson. Finite-range regularisation is
implemented to evaluate loop integrals, for reasons discussed above.
The aim is to test whether it produces a more satisfactory description
of the complete data set than the more commonly used, naive
extrapolation formulas.  A condensed version of some the work featured
here has been reported in Ref.~\cite{Allton:2005fb}.

The next section summarises the finite-range regularised forms for the
self-energy of the $\rho$ meson in the case of pQQCD.
Section~\ref{sec:cppacs} discusses the data used from 
the CP-PACS Collaboration~\cite{cppacs}.
We then give details of the chiral fits in Sec.~\ref{sec:fits}.
Finally,
Sec.~\ref{sec:expt} reports the experimental determination of the
$\rho$ meson mass at the physical point.  



\section{Self-Energies for the Partially Quenched Analysis}
\label{sec:pq}

Theoretical calculations of dynamical-fermion QCD provide an
opportunity to explore the properties of QCD in an expansive manner.
The idea is that the sea-quark masses (considered in creating the
gauge fields of the QCD vacuum) and valence quark masses (associated
with operators acting on the QCD vacuum) need not match.  Such
simulation results are commonly referred to as partially quenched
calculations.  Unlike quenched QCD, which connects to full QCD only in
the heavy quark limit, partially quenched QCD is not an approximation.
The chiral coefficients of terms in the chiral expansion (such as the
axial couplings of the $\pi$ and $\eta'$) are the same as in full QCD.
Hence, the results of partially quenched QCD provide a theoretical
extension of QCD \cite{Sharpe:2001fh}. QCD, as realized in nature, is
recovered in the limit where the valence and sea masses match.

In this section we explain the form of the finite-range regularised
chiral extrapolation formula in the case of partially quenched QCD
(pQQCD) --- i.e., the case where the valence and sea quarks are not
necessarily mass degenerate.  This work extends on the early work of
Ref.~\cite{Leinweber:1993yw}
and the more recent analysis of
Ref.~\cite{adel_rho}, which studied the case of physical (full) QCD.
This discussion includes the self-energies $\Sigma_{\pi\pi}^\rho$ and
$\Sigma_{\pi\omega}^\rho$ (corresponding to Eqs.(3) and (4) in
Ref.~\cite{adel_rho}) and includes the self-energy contributions
associated with the double hairpin diagrams surviving to some
extent in pQQCD.  We restrict our attention to the case where the two
valence quarks in the vector meson are degenerate.

We introduce the following notation.
$M_{PS(V)}(\beta,\ksea;\kval^1,\kval^2)$ refers to the
pseudoscalar (vector) meson mass, with the first two arguments
referring to the gauge coupling and sea quark mass, while the last two refer to the
valence quark masses.
Throughout the paper it will be convenient to use a short hand notation:
%

%
\bea
\ba{rcl}
M^{non-deg} &=& M(\beta,\ksea;\ksea,\kval)\mathrm \nonumber \\
M^{deg} &=& M(\beta,\ksea;\kval,\kval) \nonumber \\
M^{unit} &=& M(\beta,\ksea;\ksea,\ksea). \nonumber
\ea
\eea
%

%
where the superscript {\em unit}
refers to the unitary data with $\kval^1 \equiv \kval^2 \equiv \ksea$;
{\em deg}
refers to the ``degenerate'' data, where $\kval^1 \equiv \kval^2$
and these are not necessarily equal to $\ksea$;
{\em non-deg}
refers to the non-degenerate case where $\kval^1 = \ksea \ne \kval^2$.

\begin{figure}[tbp] 
\begin{center} 
\includegraphics[angle=0,width=5cm]{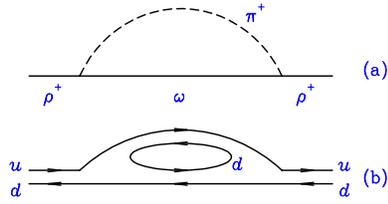}
\caption{The leading nonanalytic contribution to the chiral expansion
  of the $\rho$-meson mass.  The meson dressing (a) and its associated
  quark-flow diagram (b) are illustrated.
\label{fg:rhoPsOmega}} 
\end{center} 
\end{figure}

Derivation of the pQQCD chiral expansion for mesons can be described
via the diagrammatic method \cite{Leinweber:2002qb}, where the role of
sea-quark loops in the creation of pseudoscalar meson dressings of the
vector meson is easily observed.  Consider the simplest case of the
$\pi$-$\omega$ dressing of the $\rho$ meson depicted in
Fig.~\ref{fg:rhoPsOmega}, which gives rise to the leading nonanalytic (LNA)
contribution to the chiral expansion of the $\rho$-meson mass in full
QCD.  Here the positive charge state of the $\rho$ is selected to
simplify the derivation.
%
%

\begin{figure}[tbp] 
\begin{center} 
\includegraphics[angle=0,width=5cm]{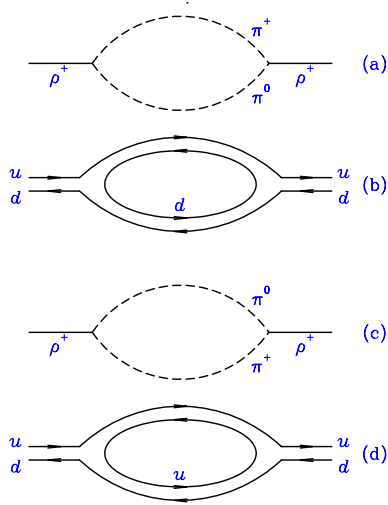}
\caption{Two-pion contributions to the positively charged $\rho$-meson
self-energy.  The quark-flow diagrams corresponding to the meson
dressings of (a) and (c) are illustrated in (b) and (d) respectively.
\label{fg:rhoPsPs}} 
\end{center} 
\end{figure}

%
The two-pion contribution to the $\rho$-meson self-energy is depicted
in Fig.~\ref{fg:rhoPsPs}.  This channel gives rise to the
next-to-leading nonanalytic (NLNA) contribution to the chiral
expansion of the $\rho$-meson mass. Importantly, this contribution
also ensures that the $\rho$ develops a finite width as the two-$\pi$
decay becomes accessible.
%
%

%
As these channels can only come about through the inclusion of a
sea-quark loop, the expressions for the pionic self-energies are as
given in Ref.~\cite{adel_rho}, but with the pion mass being
$M^{non-deg}_{PS}$, corresponding to one valence quark and one sea
quark.

In the case of the partially-quenched $\eta'$ contributions to the
chiral expansion of pQQCD, we simplify the calculation by assuming (as
in Ref.~\cite{Labrenz:1996jy}) that the Witten-Veneziano coupling is
simply a constant, $- \mu_0^2$, and take the limit $\mu_0 \rightarrow
\infty$ after summing all relevant sea quark bubble diagrams. The
philosophy is that the physical $\eta'$ mass is so large that we can
ignore $\eta'$ loops in full QCD.

\begin{figure}[tbp] 
\begin{center} 
\includegraphics[angle=0,width=5cm]{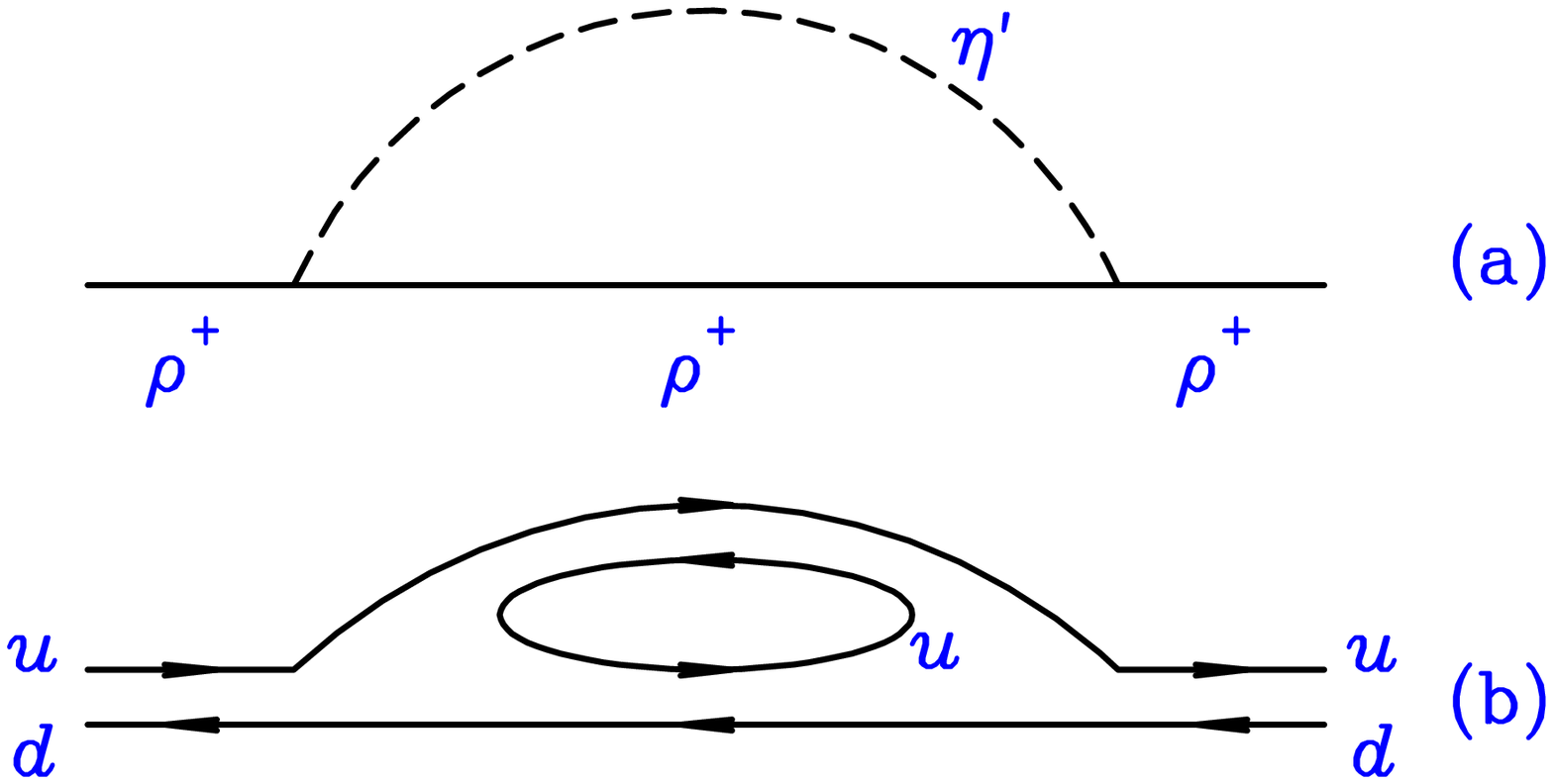}\\[0.3cm]
\includegraphics[angle=0,width=5cm]{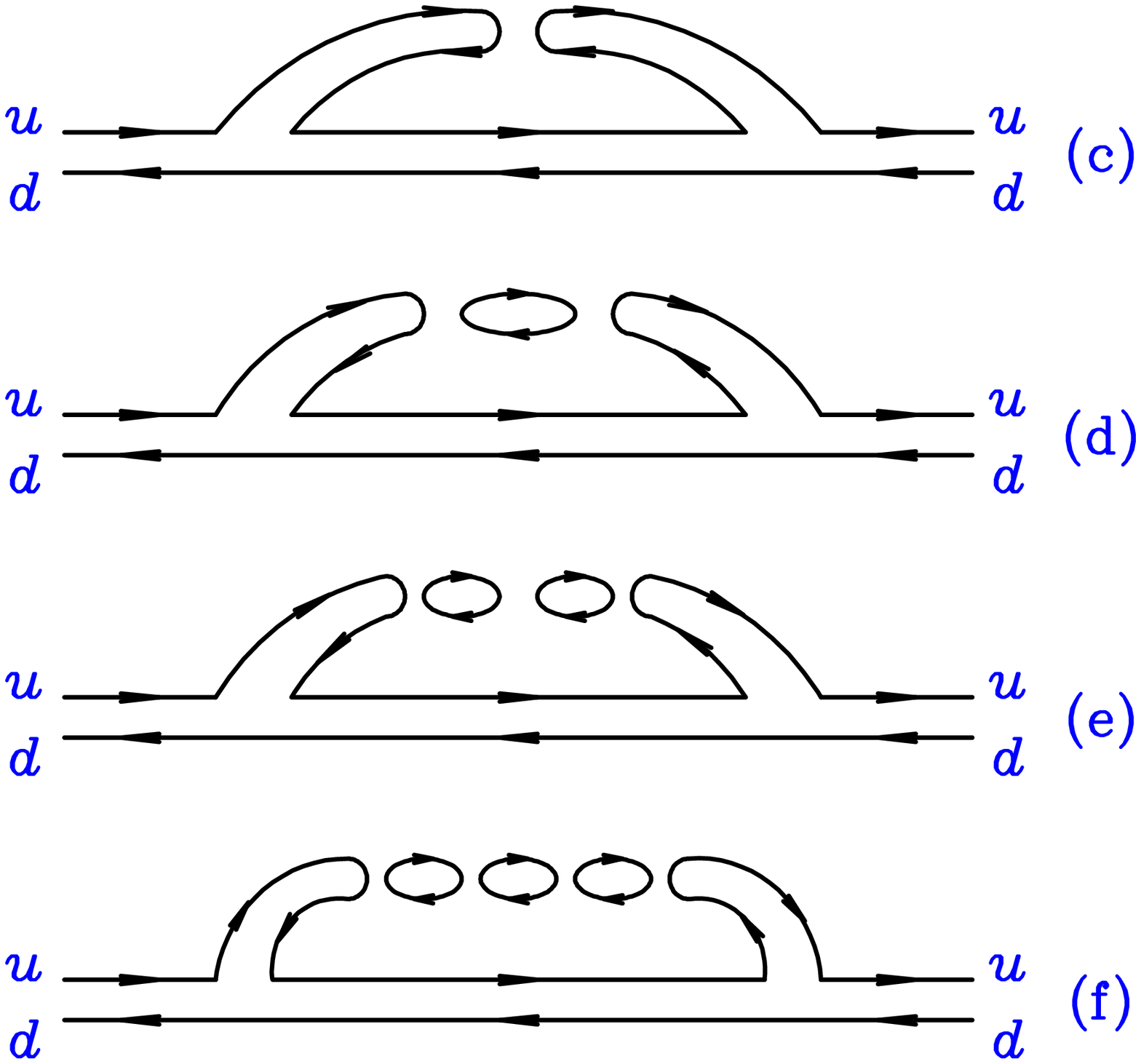}
\caption{The $\eta'$ contribution to the $\rho$-meson self energy and
  its associated quark flow diagrams in pQQCD.  While diagram (c)
  appears in quenched QCD, it is complemented by an infinite series of
  terms, the first few of which are depicted in diagrams (d) through
  (f).
\label{fg:rhoEtaVect}} 
\end{center} 
\end{figure}

The $\eta'$ contributions to the $\rho$-meson self energy are depicted in
Fig.~\ref{fg:rhoEtaVect}(a) with the associated quark flow diagrams
illustrated in Fig.~\ref{fg:rhoEtaVect}(b) through (f).  While
Fig.~\ref{fg:rhoEtaVect}(c) appears in quenched QCD, it is
complemented by an infinite series of terms, the first few of which
are depicted in Fig.~\ref{fg:rhoEtaVect}(d) through (f).  Only the sum
of Fig.~\ref{fg:rhoEtaVect}(b) through (f) and beyond raise the
$\eta'$ mass in full QCD.  
%
%

%

In partially quenched QCD it is essential to track the masses of the
pseudoscalar mesons contributing to the quark-flow diagrams of
Fig.~\ref{fg:rhoEtaVect}.  Whereas Fig.~\ref{fg:rhoEtaVect}(b)
involves a non-degenerate pseudoscalar meson,
Fig.~\ref{fg:rhoEtaVect}(c) involves degenerate pseudoscalar mesons,
and the remaining figures involve both degenerate and non-degenerate
pseudoscalar mesons.  The sum of diagrams (b) through (f) of
Fig.~\ref{fg:rhoEtaVect} and beyond generate an $\eta'$ propagator
proportional to
\begin{equation}
\frac{g_2^2}{q^2 + {M^{non-deg}_{PS}}^2} - \frac{g_2^2 \, \mu_0^2}{\left ( q^2 + {M^{deg}_{PS}}^2 \right )^2} \left [
1 - \frac{\mu_0^2}{q^2 + {M^{unit}_{PS}}^2} +
\left ( \frac{\mu_0^2}{q^2 + {M^{unit}_{PS}}^2} \right )^2 - \cdots
\right ] \, .
\label{eq:dhp}
\end{equation}
Upon summing the terms in $[ \cdots ]$, Eq.~(\ref{eq:dhp}) takes the form
\begin{equation}
\frac{g_2^2}{q^2 + {M^{non-deg}_{PS}}^2} 
- g_2^2 \frac{q^2 + {M^{unit}_{PS}}^2}{\left ( q^2 + {M^{deg}_{PS}}^2
  \right )^2} \, 
\frac{\mu_0^2}{q^2 + {M^{unit}_{PS}}^2 + \mu_0^2} \, .
\label{eq:etaPrimeFinite}
\end{equation}
We note that for equal valence- and sea-quark masses where
$M^{deg}_{PS} = M^{non-deg}_{PS} = M^{unit}_{PS}$,
Eq.~(\ref{eq:etaPrimeFinite}) describes the propagation of a heavy
$\eta'$ meson.
Upon taking $\mu_0 \to \infty$
\begin{equation}
\sigma^{\rm tot}_{\eta'\rho}  \sim  
\frac{g_2^2}{q^2 + {M^{non-deg}_{PS}}^2} 
- \frac{g_2^2 \left ( q^2 + {M^{unit}_{PS}}^2 \right )}
             {\left ( q^2 + {M^{deg}_{PS}}^2 \right )^2} \, ,
\label{eq:DHP_tot_1}
\end{equation}
By subtracting and adding $\left (q^2 + {M^{deg}_{PS}}^2 \right
)^{-1}$ to the first and second terms respectively, this can be
rearranged as
\begin{equation}
\sigma^{\rm tot}_{\eta' \rho}  \sim 
\frac{g_2^2({M^{deg}_{PS}}^2-{M^{non-deg}_{PS}}^2)}{(q^2 + {M^{non-deg}_{PS}}^2)(q^2 + {M^{deg}_{PS}}^2)} +
\frac{g_2^2 ({M^{deg}_{PS}}^2 - {M^{unit}_{PS}}^2)}{(q^2 + {M^{deg}_{PS}}^2)^2}
\label{eq:DHP_tot}
\end{equation}
which clearly vanishes when the sea and valence quark masses are equal.

\begin{figure}[tbp] 
\begin{center} 
\includegraphics[angle=0,width=5cm]{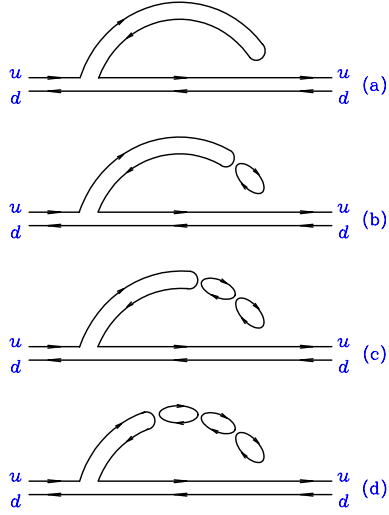}
\caption{The quenched coupling of a light $\eta'$ meson (degenerate
  with the pion) depicted in (a) is complemented by an infinite sum of
  sea-quark loop contributions in pQQCD, the first few of which are
  depicted in (b) through (d).
\label{fg:rhoEtaVert}} 
\end{center} 
\end{figure}

The quenched coupling of a light $\eta'$ meson (degenerate with the
pion) depicted in Fig.~\ref{fg:rhoEtaVert}(a) is complemented by an
infinite sum of sea-quark loop contributions in pQQCD, the first few
of which are depicted in Fig.~\ref{fg:rhoEtaVert}(b) through (d).
Using similar techniques, one can show that the sum of these graphs
leads to the propagation of a heavy $\eta'$ and upon taking $\mu_0 \to
\infty$ the contribution of Fig.~\ref{fg:rhoEtaVert} vanishes.
%
%

We note that the coupling $g_2$ introduced in Ref.~\cite{Chow:1997dw},
takes the value 0.75, where in terms of the usual $\omega \rho \pi$
coupling constant, $g_{\omega \rho \pi}$ = 16 GeV$^{-1}$, $g_2^2 =
g_{\omega \rho \pi}^2 f_\pi^2 /4$. Introducing
$f_{\rho\pi\omega}^2=\mu_\rho g_2^2$ (with $\mu_{\rho(\pi)}$ the
physical $\rho(\pi)$ mass), the total, non-trivial $\eta'$
contribution can be written in the form (following
Ref.~\cite{adel_rho} for large vector meson mass):
\begin{eqnarray}
\label{eq:sigma_dhp}
\Sigma^\rho_{\eta' \rho      }&=&\frac{f_{\rho\pi\omega}^2}{3\pi^{2} f_\pi^2}
   \int_{0}^{\infty} \frac{k^{4} \, u^{2}(k)~dk}
            {(k^{2} + (M_{PS}^{non-deg})^2)\, (k^{2} +
   (M_{PS}^{deg})^2)} \times 
            ((M_{PS}^{non-deg})^2 - (M_{PS}^{deg})^2) \nonumber \\
                         &+&\frac{f_{\rho\pi\omega}^2}{3\pi^{2} f_\pi^2}
   \int_{0}^{\infty} \frac{k^{4}\, u^{2}(k)~dk}
            {(k^{2} + (M_{PS}^{deg})^2)^2} \times
            ((M_{PS}^{unit})^2 - (M_{PS}^{deg})^2) \, ,
\end{eqnarray}
In summary, we find
%
%
\be
\label{eq:sigma_tot}
\Sigma_{TOT} = \Sigma^\rho_{\pi\pi   } +
               \Sigma^\rho_{\pi\omega} +
               \Sigma^\rho_{\eta' \rho} \, ,
\ee
%
%
where the pion self-energies are given by:
%
%
\bea 
\label{eq:sigma_pipi}
\Sigma^\rho_{\pi\pi   }&=&-\frac{f_{\rho\pi\pi}^{2}}{6\pi^{2}}
    \int_{0}^{\infty} \frac{k^{4} \, u_{\pi\pi}^{2}(k)~dk}
            {\omega_{\pi}(k)\, (\omega_{\pi}^{2}(k) - \mu_{\rho}^{2} / 4)}, \\
\label{eq:sigma_piomega}
\Sigma^\rho_{\pi\omega}&=&-\frac{f_{\rho\pi\omega}^2}{3\pi^{2}f_\pi^2}
    \int_{0}^{\infty} \frac{k^{4} \, u_{\pi\omega}^{2}(k)~dk}
            {\omega_{\pi}(k)\, ( \omega_\pi(k) + \Delta M_{\omega\rho}
    )}\, ,
\eea
%
and
\begin{equation}
\omega_{\pi}^{2}(k) = k^{2} + (M^{non-deg}_{PS})^2 \, .
\end{equation}
As the $\omega$ meson contributes via a sea-quark loop as illustrated
in Fig.~\ref{fg:rhoPsOmega}, the mass splitting between the $\rho$ and
$\omega$ is
\be
\Delta M_{\omega\rho} = M_V^{non-deg} - M_V^{deg} \, . 
\ee
We note that $(\omega_\pi(k) + \Delta M_{\omega\rho}) > 0$ for all
quark masses and nontrivial momenta considered in the lattice
analysis.  We use the values
$f_{\rho\pi\pi} = 6.028$ (obtained from Ref.~\cite{adel_rho}) and
$f_{\pi} = 92.4\mev$.
We use a standard dipole form factor, which takes the form
%
\bea
u(k)&=&\frac{\Lambda^{4}}{(\Lambda^{2}+k^{2})^2} \, , \nonumber \\
u_{\pi\omega}(k)&=&u(k) \, , \nonumber \\
u_{\pi\pi}(k)&=&u(k)\, u^{-1} ( \sqrt{\mu_\rho^2 / 4 - \mu_\pi^2}) \, , \nonumber
\eea
%
where the second factor in $u_{\pi\pi}(k)$ ensures the correct on-shell
normalization condition.

To account for finite volume artefacts, 
the self-energy equations are discretised so that only those momenta
allowed on the lattice appear \cite{adel_rho,Young:2002cj}:
%
\be
4 \pi \int_{0}^{\infty}k^2dk = \int d^3k \approx
\frac{1}{V}\left(\frac{2 \pi}{a}\right)^3 \sum_{k_x,k_y,k_z} \, ,
\label{eq:lat_int}
\ee
with
\be
k_{x,y,z} = \frac{2\pi(i,j,k)}{a N_{x,y,z}} \, .
\ee
%
%
The purpose of the finite-range regulator is to regularise the theory
as $k_x$, $k_y$, $k_z$ tend to infinity. Indeed, once any one of
$k_x$, $k_y$ or $k_z$ is greater than $\sim 10\Lambda$ the
contribution to the integral is negligible and thereby ensuring
convergence of the summation.  Hence, we would like the highest
momentum in each direction to be just over $10\Lambda$.  For practical
calculation, we therefore use the following to calculate the maxima
and minima for i, j, k:
%
%
\bea
    (i,j,k)_{max} &=& ~~\left
    [\frac{10\Lambda~a}{2\pi}~N_{(x,y,z)}\right ] + 1, \nonumber \\
    (i,j,k)_{min} &=&  -\left [\frac{10\Lambda~a}{2\pi}~N_{(x,y,z)}
    \right ] - 1, \nonumber
\eea
%
%
where $[\ldots]$ denotes the integer part.

\begin{figure}[*htbp] 
\begin{center} 
\includegraphics[angle=0, width=0.85\textwidth]{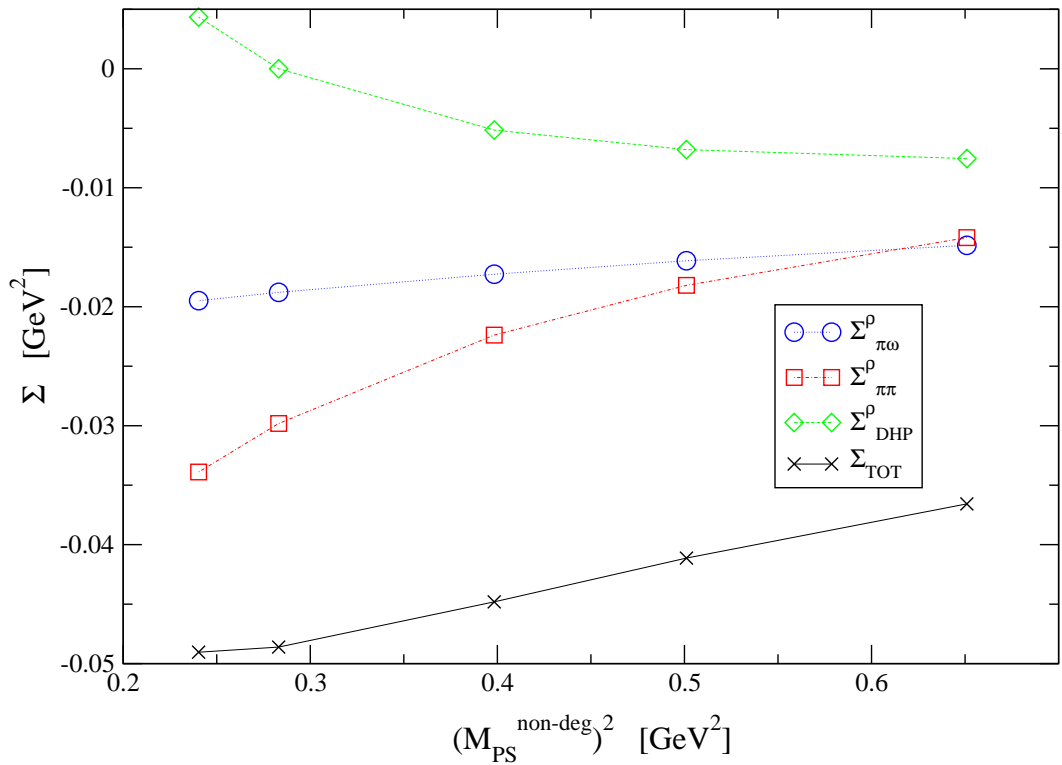}
\caption{The self-energy contributions 
(see Eqs. \ref{eq:sigma_pipi}, \ref{eq:sigma_piomega} \& \ref{eq:sigma_dhp})
versus $M_{PS}^{non-deg}$ data for the ensemble
$(\beta,\ksea) = (2.10,0.1382)$ using $\Lambda = 630$ MeV.
\label{fg:self_energy}} 
\end{center} 
\end{figure}

We study a range of values of $\Lambda$, starting with the value,
$\Lambda_{\pi\omega} = 630$ MeV, used in Ref.~\cite{adel_rho}.
{}Figure \ref{fg:self_energy} shows the various self-energy
contributions, $\Sigma^\rho_{\pi\pi}, \Sigma^\rho_{\pi\omega}$ and
$\Sigma^\rho_{\eta' \rho}$ as a function of $M_{PS}^{non-deg}$ (see
Eqs.~(\ref{eq:sigma_pipi}), (\ref{eq:sigma_piomega}) and
(\ref{eq:sigma_dhp}) ) for the representative
$(\beta,\ksea)=(2.10,0.1382)$ dataset.  In Sec.~\ref{sec:global} we
perform a highly constrained fit to a large ``global'' dataset, and 
this enables us
to determine a best value of $\Lambda$ which minimises the
global $\chi^2$.
%

%
%




\section{Overview of CP-PACS Data}
\label{sec:cppacs}

In Ref.~\cite{cppacs}, the CP-PACS collaboration published meson spectrum
data from dynamical simulations for mean-field improved Wilson
fermions with improved gluons at four different $\beta$ values. For
each value of $\beta$, ensembles were generated for four different 
values of $\ksea$ -- 
giving a total of 16 independent ensembles.
Table~\ref{tb:lat} summarises the lattice parameters used. 
The pseudoscalar- to vector-meson mass ratio
$M^{unit}_{PS}/M^{unit}_{V}$ (which gives a measure 
of the mass of the sea quarks used) varies from 0.55 to 0.80
and the lattice spacing, $a$, from around 0.09 to 0.29 fm.
{}For each of the 16 ensembles there are five $\kval$ values considered
\cite{cppacs}. Thus there are a total of 80 $(M_V^{deg},M_{PS}^{deg})$
data points available for analysis.
In our study we consider the two cases where the lattice spacing is
set using either $r_0$ \cite{sommer} or the string tension $\sigma$.

%

\begin{table}[*htbp]
\begin{center}
\begin{ruledtabular}
\begin{tabular}{ccclll}
$\beta$ & $\kappa_{sea}$ & Volume & \multicolumn{1}{c}{$M_{PS}^{unit}/M_V^{unit}$} & \multicolumn{1}{c}{$a_{r_0}$ [fm]} & \multicolumn{1}{c}{$a_{\sigma}$ [fm]} \\
\hline
%
%
  1.80 & 0.1409 & $12^3 \times 24$ &  0.8067\err{ 9}{ 9} & 0.286\err{ 6}{ 6}  & 0.288\err{ 3}{ 3}   \\
  1.80 & 0.1430 & $12^3 \times 24$ &  0.7526\err{16}{15} & 0.272\err{ 2}{ 2}  & 0.280\err{ 4}{ 5}   \\
  1.80 & 0.1445 & $12^3 \times 24$ &  0.694\err{ 2}{ 2}  & 0.258\err{ 4}{ 4}  & 0.269\err{ 2}{ 3}   \\
  1.80 & 0.1464 & $12^3 \times 24$ &  0.547\err{ 4}{ 4}  & 0.237\err{ 4}{ 4}  & 0.248\err{ 2}{ 3}   \\
\hline
  1.95 & 0.1375 & $16^3 \times 32$ &  0.8045\err{11}{11} & 0.196\err{ 4}{ 4}  & 0.2044\err{10}{12}   \\
  1.95 & 0.1390 & $16^3 \times 32$ &  0.752\err{ 2}{ 2}  & 0.185\err{ 3}{ 3}  & 0.1934\err{14}{15}   \\
  1.95 & 0.1400 & $16^3 \times 32$ &  0.690\err{ 2}{ 2}  & 0.174\err{ 2}{ 2}  & 0.1812\err{12}{12}   \\
  1.95 & 0.1410 & $16^3 \times 32$ &  0.582\err{ 3}{ 3}  & 0.163\err{ 2}{ 2}  & 0.1699\err{13}{15}   \\
\hline
  2.10 & 0.1357 & $24^3 \times 48$ &  0.806\err{ 2}{ 2}  & 0.1275\err{ 5}{ 5} & 0.1342\err{ 8}{ 8}   \\
  2.10 & 0.1367 & $24^3 \times 48$ &  0.755\err{ 2}{ 2}  & 0.1203\err{ 4}{ 5} & 0.1254\err{ 8}{ 8}   \\
  2.10 & 0.1374 & $24^3 \times 48$ &  0.691\err{ 3}{ 3}  & 0.1157\err{ 4}{ 4} & 0.1203\err{ 6}{ 6}   \\
  2.10 & 0.1382 & $24^3 \times 48$ &  0.576\err{ 3}{ 4}  & 0.1093\err{ 3}{ 3} & 0.1129\err{ 4}{ 5}   \\
\hline
  2.20 & 0.1351 & $24^3 \times 48$ &  0.799\err{ 3}{ 3}  & 0.0997\err{ 4}{ 5} & 0.10503\err{15}{15}  \\
  2.20 & 0.1358 & $24^3 \times 48$ &  0.753\err{ 4}{ 4}  & 0.0966\err{ 4}{ 4} & 0.1013\err{ 3}{ 2}   \\
  2.20 & 0.1363 & $24^3 \times 48$ &  0.705\err{ 6}{ 6}  & 0.0936\err{ 4}{ 4} & 0.0978\err{ 3}{ 3}   \\
  2.20 & 0.1368 & $24^3 \times 48$ &  0.632\err{ 8}{ 8}  & 0.0906\err{ 4}{ 4} & 0.0949\err{ 2}{ 2}   \\
\end{tabular} 
\end{ruledtabular}
\end{center} 
\caption{\small The lattice parameters of the CP-PACS simulation used
in this data analysis, taken from Ref.~\cite{cppacs}.  The superscript
{\em unit} refers to the unitary data (i.e., where $\kval^1 \equiv
\kval^2 \equiv \ksea$).  Note that the errors reported in this table
are obtained with our bootstrap ensembles (see text).
\label{tb:lat}} 
\end{table} 
%

%

In the absence of the full set of original CP-PACS data, 
we generate 1000 bootstrap clusters for all $M_{PS}$ and $M_V$ data
using a Gaussian distribution whose central value and full width 
half maximum (FWHM) are the
same as the central values and errors published in the Table XXI
of Ref.~\cite{cppacs}. Of course, our  
errors are {\em totally uncorrelated} throughout -- i.e., each
$M_V(\beta,\ksea;\kval^1,\kval^2)$ bootstrap cluster is uncorrelated
with the corresponding $M_{PS}(\beta,\ksea;\kval^1,\kval^2)$ bootstrap
cluster. Furthermore, the $M(\beta,\ksea;\kval^1,\kval^2)$ data is
not correlated with the $M(\beta',\ksea';\kval^1,\kval^2)$ data, and
nor is the \mbox{$M(\beta,\ksea;\kval^1,\kval^2)$} data 
correlated with that for $M(\beta,\ksea;\kval^{1'},\kval^{2'})$.

We expect therefore that the statistical errors of our final results will be
overestimates of the true error, since we have not benefited from the
partial cancellation of statistical errors which occurs when
combining correlated data. We can obtain a rough estimate of the
increase in our errors due to the fact that we don't maintain
correlations as follows. The ratio $M_{PS}^{unit}/M_V^{unit}$
listed in Table~\ref{tb:lat} is obtained from our bootstrap
data. Comparing this with the $M_{PS}^{unit}/M_V^{unit}$ data
in Table XXI of Ref.~\cite{cppacs} (which benefits from the
cancellation of correlations), we see that a very rough estimate of the 
effect of ignoring correlations is to increase the errors by $\sim $20\%.
We expect that a similar increase in errors will apply to other
quantities.

\begin{figure}[*htbp] 
\begin{center} 
\includegraphics[angle=0, width=0.85\textwidth]{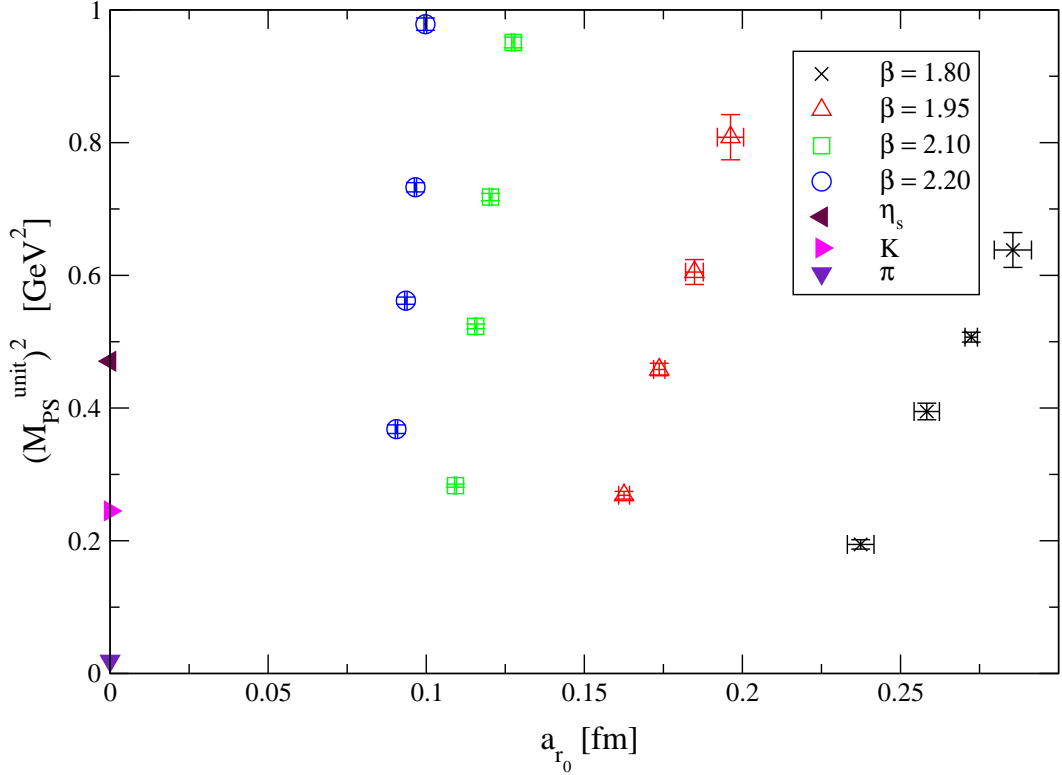}
\caption{The range of sea quark mass
$(M_{PS}^{unit})^2$ and lattice spacing, $a_{r_0}$, covered by the
\mbox{CP-PACS} data displayed in Table~\ref{tb:lat}.
$(M_{PS}^{unit})^2$ is the pseudoscalar meson mass squared
at the unitary point; i.e., where $\kval \equiv \ksea$.
The experimental points for the $\pi, K$ and ``$\eta_s$'' mesons
are also shown for reference.
\label{fg:a_r0_vs_mps2}} 
\end{center} 
\end{figure}

The lattice spacings $a_{r_0,\sigma}$ were obtained from 
Table XII of Ref.~\cite{cppacs} using $r_0 = 0.49$ fm
and $\sqrt{\sigma} = 440$ MeV.
Again we generated 1000 bootstrap clusters with a Gaussian distribution,  
as in the meson mass data.
{}Figure \ref{fg:a_r0_vs_mps2} shows the unitary (i.e., $\kval^1 \equiv
\kval^2 \equiv \ksea$) pseudoscalar mass plotted against the lattice
spacing, $a_{r_0}$, for the 16 ensembles in Table~\ref{tb:lat}. (Note
that $(M_{PS}^{unit})^2$ is a direct measure of the sea quark mass
since, from PCAC, $(M_{PS}^{unit})^2 \propto \msea$.)
Also shown for reference are the physical
pseudoscalar mesons $\pi, K$ and $``\eta_s"$.
Note the large range of both the lattice spacing $a_{r_0}$ and $\msea$
in the simulations, and that the lattice spacing is primarily
determined by the $\beta$ value, rather than the value of $\msea$.

The physical volume for these 16 ensembles is $La \approx 2.5$ fm for
the $\beta=1.80, 1.95$ and $2.10$ cases, but the $\beta=2.20$ ensemble had a
slightly smaller physical volume.  The associated finite volume
effects are incorporated through evaluating the chiral loops by
explicitly summing the discrete pion momenta allowed on the lattice.
We treat all 16 ensembles on an equal footing.  The finite-volume
effects are corrected when making contact with the physical
observables by evaluating the chiral loop integrals with 
continuous loop momenta.

The action used in Ref.~\cite{cppacs} is mean-field improved, rather
than non-perturbatively improved and will therefore have some residual
lattice systematic errors of ${\cal O}(a)$ \cite{scaling}.  We fit the
data assuming both ${\cal O}(a)$ and ${\cal O}(a^2)$ effects in
sections \ref{sec:individual} and \ref{sec:global}, and are thus able
to obtain continuum predictions.  Our empirical analysis suggests that
nonanalytic terms generated in a dual expansion of both $a$ and $m_q$
\cite{Bar:2004xp,Rupak:2002sm,Bar:2003mh,Aoki:2003yv,Beane:2003xv,Grigoryan:2005zj}
are either small or can be absorbed at present
into the ${\cal O}(a)$ and ${\cal O}(a^2)$ effects considered here.
%


%
%

%

 

\section{Fitting Analysis} 
\label{sec:fits}


\subsection{Summary of Analysis Techniques}
\label{sec:ourmethod}

Our chiral extrapolation approach is based upon converting all masses
into physical units prior to any extrapolation being performed. An
alternative approach would be to extrapolate dimensionless masses (in
lattice units) \cite{cppacs}.  However, using physical units offers
the following key  
advantages:
%

%
\begin{itemize}
\item The data from different ensembles can be combined in a global
fit. If the masses were left in dimensionless units, there would be no
possibility of combining data at different lattice spacings together
into such a global fit.
\item Dimensionful mass predictions from lattice simulations are
effectively mass {\em ratios}, and hence one would expect some of the
systematic (and statistical) errors to cancel. That is, $M^{dimful} = M^\#
\times a^{-1} \equiv M^\# / M_\Omega^\# \times M_\Omega^{expt}$, where
$\Omega$ is the quantity used to set the lattice spacing, $a$,
the superscripts \#, $expt$ refer to the lattice mass estimate
and the experimental value, respectively.
\item Lattice data are not in the regime where the coefficients of
nonanalytic terms in the chiral expansion can be reliably
constrained. These dimensionful quantities must therefore be fixed to
their phenomenological values.
\end{itemize}

%
In this paper we use two different quantities for setting the scale,
$r_0$ and $\sigma$ -- although we have a preference, as explained 
in Sec.~\ref{sec:global}, for $r_0$.
Table~\ref{tb:lat} lists values for $a_{r_0}$ and $a_\sigma$.

In our chiral extrapolations we use two basic fitting functions,
the finite-range regularisation method 
(hereafter referred to as the ``Adelaide''
method)
%

\bea\label{eq:adel}
\sqrt{(M_V^{deg})^{2} - \Sigma_{TOT}} &=&
a_0 + a_2 (M_{PS}^{deg})^2 + a_4 (M_{PS}^{deg})^4 + a_6 (M_{PS}^{deg})^6,
\eea
%

%
\noindent where $\Sigma_{TOT}$ is calculated using Eq.~(\ref{eq:sigma_tot}),
and a naive polynomial fit
%
\be\label{eq:naive}
M_V^{deg} =
a_0 + a_2 (M_{PS}^{deg})^2 + a_4 (M_{PS}^{deg})^4 + a_6 (M_{PS}^{deg})^6.
\ee
%
%
In each case we refer to these fits as ``cubic'' since they include
cubic terms in the chiral expansion of $m_q \propto (M_{PS})^2$.
We also perform fits with the coefficient $a_6$ set to zero
in Eqs.(\ref{eq:adel}) and (\ref{eq:naive}), referring to these as ``quadratic''.
It is worth noting that the dominant functional form of $M_V$ with
$(M_{PS}^{deg})^2$ is linear (see, for example,
Fig. \ref{fg:second_lightest}, where the LHS of Eqs.~(\ref{eq:adel})
and (\ref{eq:naive}) are referred to as $M_V^{\rm sub}$).
We exploit this fact in the fitting functions given above.
In particular, this is why the Adelaide fit uses 
$\sqrt{(M_V^{deg})^{2} - \Sigma_{TOT}}$ on the LHS rather than
$(M_V^{deg})^{2} - \Sigma_{TOT}$ which would, a priori, be an equally
valid chiral expansion.
Thus, with the above functional forms, we expect the coefficients,
$a_n$, to be
small for $n \ge 4$ and this is, in fact, what we find.

In the following, we first use Eqs.~(\ref{eq:adel}) and (\ref{eq:naive})
to fit the 16 ensembles in Table~\ref{tb:lat} {\em separately}.
We then turn to a {\em global} fit, where all 16 ensembles are combined
in one fitting function.



\subsection{Individual ensemble fits}
\label{sec:individual}

We begin our analysis by fitting the meson spectrum of each of the 16 ensembles
listed in Table~\ref{tb:lat} {\em separately}. In this section we
use $r_0$ to set the scale and select a value of
$\Lambda = 650$ MeV (see Sec.~\ref{sec:global}).
We use both the Adelaide, Eq.~(\ref{eq:adel}), and naive,
Eq.~(\ref{eq:naive}), 
fitting functions and restrict our attention to quadratic ($a_6 \equiv 0$)
chiral fits, since there are only five $(M_V^{deg},M_{PS}^{deg})$
data points available for each ensemble.
The coefficients, $a_{0,2,4}$,  
obtained by fitting $M_V$ against
$M_{PS}$ with both the naive (Eq.~(\ref{eq:naive})) and Adelaide
(Eq.~(\ref{eq:adel})) fitting functions are listed in
Table~\ref{tb:individual}.
We see that the $a_4$ coefficients are
both small and generally poorly determined, confirming our decision to
fit to the quadratic, rather than the cubic, chiral extrapolation form.
We note also that there is some agreement between the
naive and Adelaide $a_{0,2}$ coefficients, although their
variation with $\ksea$ tends to be different.

%

\begin{table}[*htbp]
\begin{center}
\begin{ruledtabular}
\begin{tabular}{ccllllll}
  $\beta$ & $\kappa_{sea}$ & \multicolumn{1}{c}{$a^{naive}_{0}$} & \multicolumn{1}{c}{$a^{adel}_{0}$} & 
\multicolumn{1}{c}{$a^{naive}_{2}$} & \multicolumn{1}{c}{$a^{adel}_{2}$} & \multicolumn{1}{c}{$a^{naive}_{4}$} & 
\multicolumn{1}{c}{$a^{adel}_{4}$}  \\
  & & \multicolumn{1}{c}{[GeV]} & \multicolumn{1}{c}{[GeV]} & \multicolumn{1}{c}{[GeV$^{-1}$]} & 
      \multicolumn{1}{c}{[GeV$^{-1}$]} & \multicolumn{1}{c}{[GeV$^{-3}$]} & \multicolumn{1}{c}{[GeV$^{-3}$]} \\
\hline
   1.80 & 0.1409 & 0.701\err{14}{22} & 0.70\err{ 2}{ 2}  & 0.46\err{ 7}{ 3} &  0.54\err{ 5}{ 5} & -0.01\err{ 3}{ 7} &  -0.09\err{ 5}{ 5}  \\
   1.80 & 0.1430 & 0.712\err{14}{13} & 0.724\err{14}{13} & 0.48\err{ 6}{ 6} &  0.51\err{ 5}{ 6} & -0.04\err{ 6}{ 6} &  -0.08\err{ 6}{ 6}  \\
   1.80 & 0.1445 & 0.73\err{ 2}{ 2}  & 0.756\err{14}{15} & 0.43\err{ 5}{ 5} &  0.44\err{ 5}{ 5} &  0.01\err{ 5}{ 5} &  -0.01\err{ 5}{ 5}  \\
   1.80 & 0.1464 & 0.72\err{ 2}{ 2}  & 0.769\err{13}{15} & 0.49\err{ 5}{ 5} &  0.43\err{ 5}{ 5} & -0.02\err{ 6}{ 6} &  0.007\err{59}{58}  \\
\hline
   1.95 & 0.1375 & 0.76\err{ 2}{ 2}  & 0.75\err{ 2}{ 2}  & 0.49\err{ 4}{ 4} &  0.53\err{ 4}{ 4} & -0.05\err{ 4}{ 3} &  -0.08\err{ 3}{ 3}  \\
   1.95 & 0.1390 & 0.76\err{ 2}{ 2}  & 0.772\err{17}{15} & 0.47\err{ 4}{ 4} &  0.49\err{ 4}{ 4} & -0.03\err{ 4}{ 3} &  -0.05\err{ 4}{ 4}  \\
   1.95 & 0.1400 & 0.785\err{12}{12} & 0.803\err{11}{11} & 0.43\err{ 4}{ 4} &  0.44\err{ 4}{ 4} & -0.01\err{ 3}{ 3} &  -0.02\err{ 3}{ 3}  \\
   1.95 & 0.1410 & 0.766\err{13}{15} & 0.799\err{13}{14} & 0.48\err{ 5}{ 4} &  0.45\err{ 5}{ 4} & -0.03\err{ 4}{ 4} &  -0.03\err{ 3}{ 4}  \\
\hline
   2.10 & 0.1357 & 0.829\err{14}{14} & 0.820\err{14}{14} & 0.42\err{ 5}{ 4} &  0.46\err{ 5}{ 4} & -0.02\err{ 3}{ 4} &  -0.05\err{ 3}{ 4}  \\
   2.10 & 0.1367 & 0.794\err{11}{10} & 0.797\err{11}{10} & 0.50\err{ 3}{ 3} &  0.53\err{ 3}{ 3} & -0.06\err{ 3}{ 3} &  -0.08\err{ 3}{ 2}  \\
   2.10 & 0.1374 & 0.807\err{13}{14} & 0.822\err{13}{14} & 0.48\err{ 4}{ 4} &  0.49\err{ 4}{ 4} & -0.05\err{ 3}{ 3} &  -0.06\err{ 3}{ 3}  \\
   2.10 & 0.1382 & 0.781\err{10}{ 9} & 0.814\err{10}{ 9} & 0.53\err{ 3}{ 3} &  0.50\err{ 3}{ 3} & -0.08\err{ 2}{ 2} &  -0.07\err{ 2}{ 2}  \\
\hline
   2.20 & 0.1351 & 0.84\err{ 3}{ 3}  & 0.84\err{ 3}{ 3}  & 0.43\err{ 8}{ 8} &  0.46\err{ 8}{ 8} & -0.02\err{ 6}{ 6} &  -0.04\err{ 6}{ 6}  \\
   2.20 & 0.1358 & 0.83\err{ 2}{ 2}  & 0.84\err{ 2}{ 2}  & 0.44\err{ 7}{ 7} &  0.46\err{ 7}{ 7} & -0.03\err{ 5}{ 5} &  -0.05\err{ 5}{ 5}  \\
   2.20 & 0.1363 & 0.80\err{ 3}{ 3}  & 0.81\err{ 3}{ 3}  & 0.51\err{ 8}{ 8} &  0.52\err{ 8}{ 8} & -0.07\err{ 6}{ 6} &  -0.08\err{ 6}{ 6}  \\
   2.20 & 0.1368 & 0.78\err{ 2}{ 2}  & 0.80\err{ 2}{ 2}  & 0.52\err{ 8}{ 8} &  0.51\err{ 7}{ 8} & -0.06\err{ 6}{ 6} &  -0.06\err{ 6}{ 6}  \\
\end{tabular} 
\end{ruledtabular}
\end{center} 
\caption{The coefficients obtained from fitting $M_V$ data against
$M_{PS}^2$ using both the naive and Adelaide fits
(i.e., Eqs.~(\ref{eq:naive}) and (\ref{eq:adel}), respectively) for each of the
16 ensembles listed in Table~\ref{tb:lat}.
As discussed in the text we restrict these fits to quadratic rather
than cubic chiral functions (i.e., $a_6 \equiv 0$).
The scale was set using $r_0$.
\label{tb:individual}}
\end{table}
%

%

\begin{figure}[*htbp]
\begin{center}
\includegraphics[angle=0, width=0.85\textwidth]{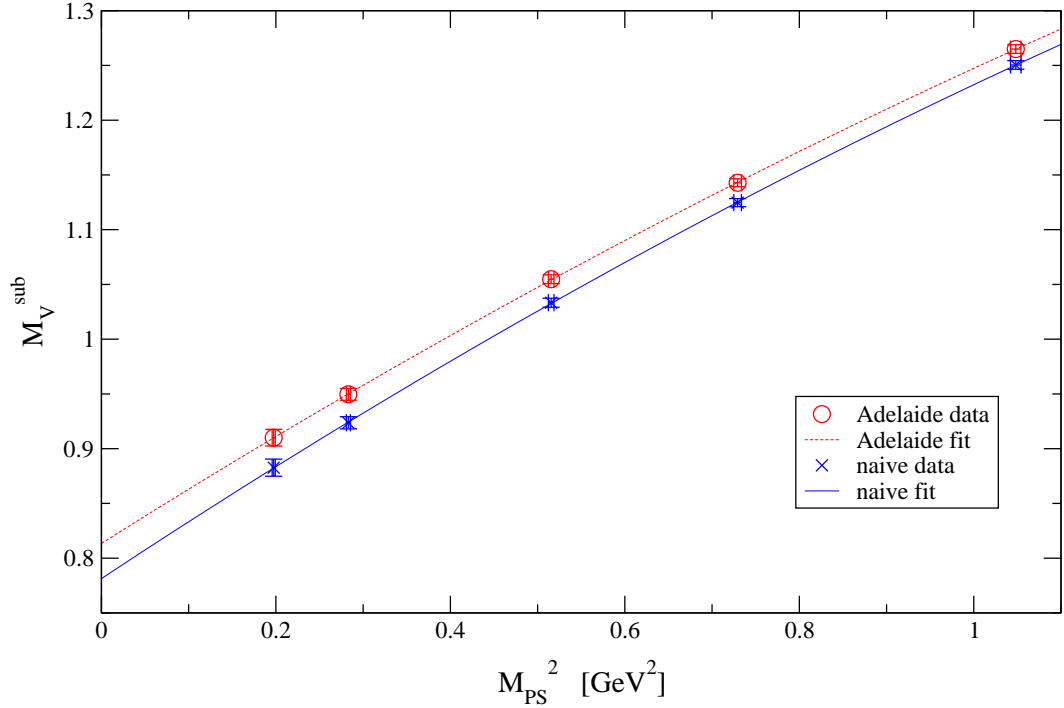}
\caption
{A plot of $M_V^{Sub}$ versus $M_{PS}$ data for the ensemble
$(\beta,\ksea) = (2.10,0.1382)$ together with the results of the
quadratic Adelaide, Eq.~(\ref{eq:adel}), and naive,
Eq.~(\ref{eq:naive}), fits.
$M_V^{Sub}$ is defined as $M_V^{Sub} = 
\sqrt{(M_V^{deg})^{2} - \Sigma_{TOT}}$
for the Adelaide fit (i.e., the LHS of Eq.~\ref{eq:adel} - note that  
$\Sigma_{TOT}$ is negative), and
$M_V^{Sub} = M_V^{deg}$
for the naive fit (i.e., the LHS of Eq.~(\ref{eq:naive})).
\label{fg:second_lightest}}
\end{center}
\end{figure}

{}Figure~\ref{fg:second_lightest} shows the results of these fits for the
$(\beta,\ksea)=(2.10,0.1382)$ ensemble, which is a good representative of all
of them.  The scale is set from $r_0$, which is our preferred method (see
Sec.~\ref{sec:global}). Note
that this ensemble's $(a,\msea)$ coordinates are closest to the
physical point $(a,\msea) = (0,m_{u,d})$ for those ensembles with $La \approx
2.5\,$fm  -- see Fig.~\ref{fg:a_r0_vs_mps2}.
%

%
%

\begin{figure}[*htbp]
\begin{center}
\includegraphics[angle=0, width=0.85\textwidth]{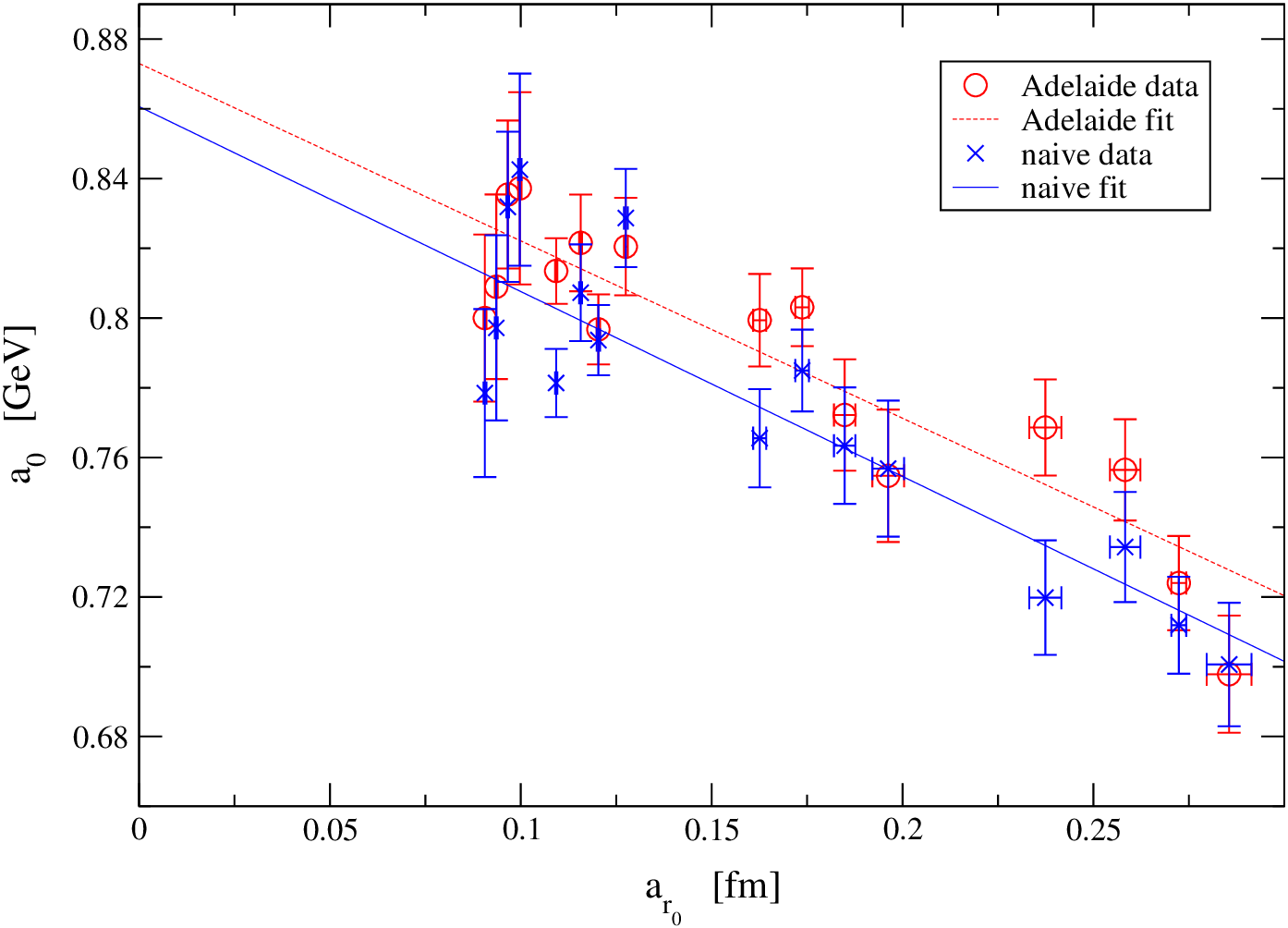}
\caption
{A continuum extrapolation of the $a_0$ coefficient obtained from both
the Adelaide and naive fits using Eq.~(\ref{eq:a02_cont}).
\label{fg:a0_cont}}
\end{center}
\end{figure}

\begin{figure}[*htbp]
\begin{center}
\includegraphics[angle=0, width=0.85\textwidth]{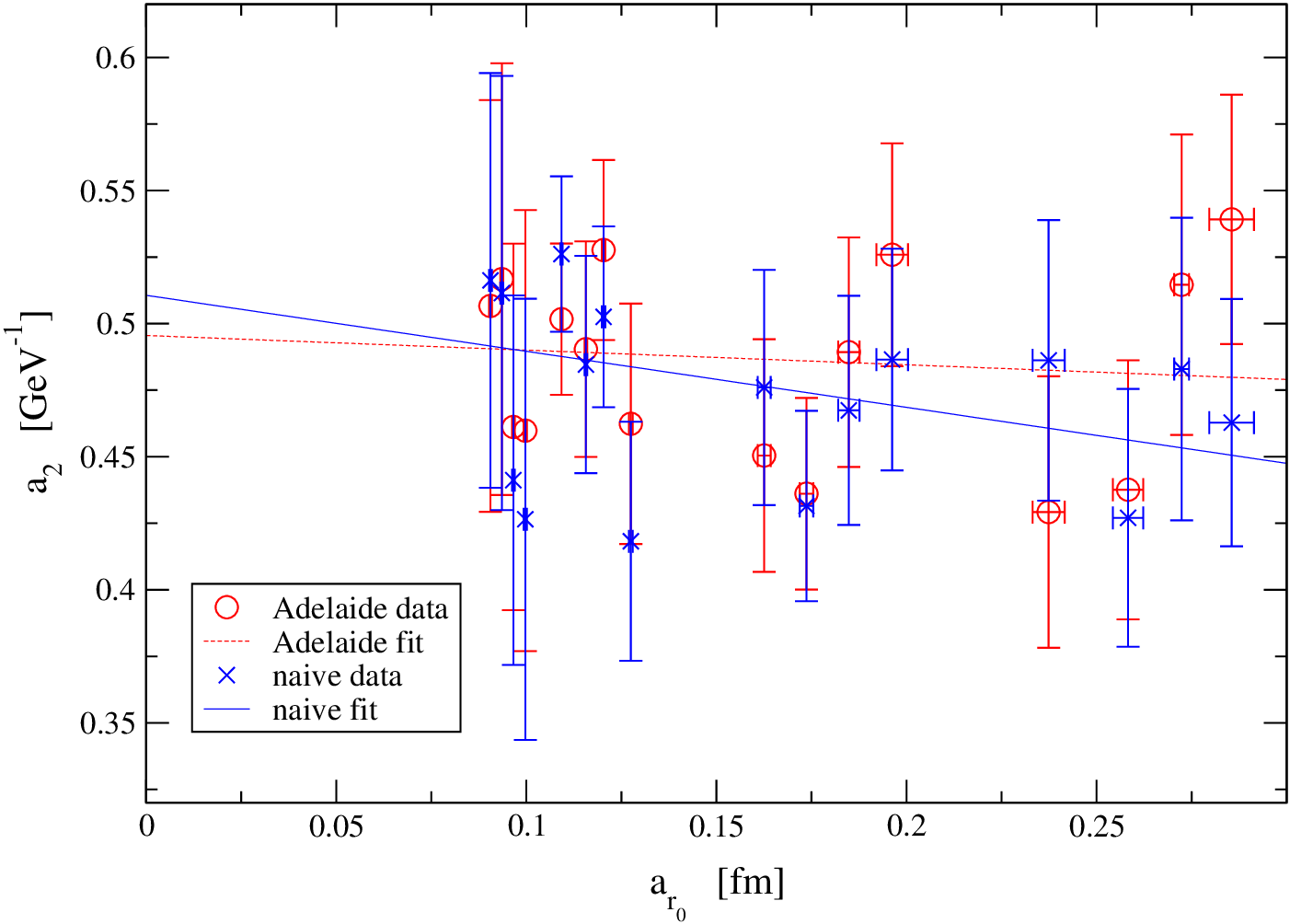}
\caption
{A continuum extrapolation of the $a_2$ coefficient obtained from both
the Adelaide and naive fits using Eq.~(\ref{eq:a02_cont}).
\label{fg:a2_cont}}
\end{center}
\end{figure}

%
The values of $a_{0,2}$ in Table~\ref{tb:individual} hint at a
systematic variation of $a_{0,2}$ with $a_{r_0}$. To check this we
plot $a_0$ and $a_2$ against $a_{r_0}$ (for both the linear and
Adelaide fits) in {}Figs.~\ref{fg:a0_cont} and \ref{fg:a2_cont}.
This motivates a continuum extrapolation of the form
%

\be\label{eq:a02_cont}
a_{0,2} = a_{0,2}^{cont} + X^{individual}_{0,2} \;a_{r_0} \, .  
\ee

%
%
%
%
%
%
%
The results of these fits are displayed in Table~\ref{tb:a02_cont}.
From the values of $X^{individual}_{0,2}$ in Table~\ref{tb:a02_cont}, we can
confirm statistically significant ${\cal O}(a)$ effects in $a_0$ but
not in $a_2$.
This will be important in determining our fit procedure in Sec.~\ref{sec:global}.

%
%
\begin{table}[*htbp]
\begin{center}
\begin{tabular}{l|ccc|ccc}
\hline\hline
  & $a_{0}^{cont.}$ & $X^{individual}_{0}$ & $\chi^{2}_{0}/d.o.f.$ &
    $a_{2}^{cont.}$ & $X^{individual}_{2}$ &  $\chi^{2}_{2}/d.o.f.$ \\
  & [GeV]           & [GeV/fm]             &                       &
    [GeV$^{-1}$]    & [GeV$^{-1}$/fm] & \\
\hline
%
%
Naive-fit  & 0.861\err{11}{ 9} & -0.53\err{ 5}{ 7} & 21 / 14 & 0.51\err{ 3}{ 4} & -0.21\err{23}{15} & 8 / 14 \\
Adelaide-fit & 0.873\err{10}{10} & -0.51\err{ 5}{ 6} & 16 / 14 & 0.50\err{ 3}{ 3} & -0.06\err{19}{18} & 10 / 14 \\
\hline\hline
\end{tabular}
\end{center}
\caption{The coefficients obtained from the continuum extrapolation of
both the naive and Adelaide fits, using the values of $a_{0,2}$ given in Table
\ref{tb:individual} with Eq.~(\ref{eq:a02_cont}).
\label{tb:a02_cont}}
\end{table}
%

%



\subsection{Global fits}
\label{sec:global}

\begin{figure}[*htbp]
\begin{center}
\includegraphics[angle=0, width=0.85\textwidth]{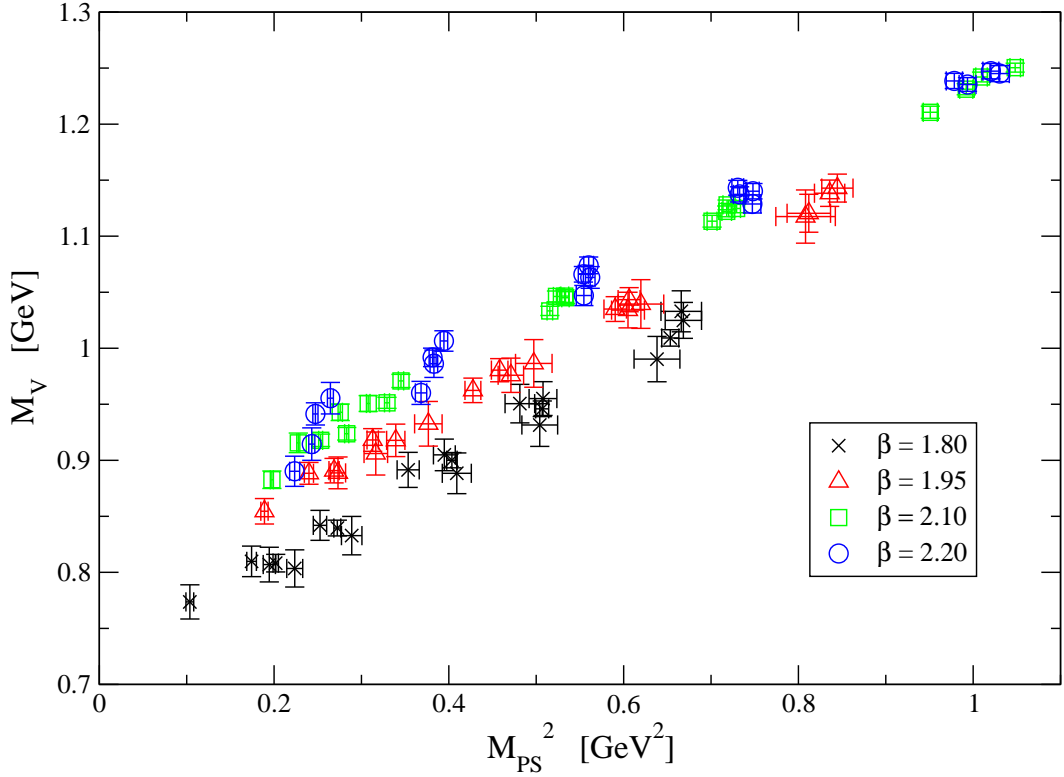}
\caption
{Illustration of the 80 lattice data points of the degenerate
\mbox{CP-PACS} data set, with the physical scale of a$_{r_0}$.
\label{fg:global}}
\end{center}
\end{figure}

We now turn our attention to an analysis of the whole (degenerate)
data set. Figure~\ref{fg:global} shows all of the degenerate CP-PACS
data with the physical scale set using $r_0$.
Since there are 16 ensembles with five $(M_V^{deg},M_{PS}^{deg})$
values in each (see Sec.~\ref{sec:cppacs}), this global fit
contains 80 data points.
We expect that this large number of data points should produce a highly
constrained set of fit parameters, $a_{0,2,\ldots}$.
%

%
%

%
Referring to the coefficients listed in Table~\ref{tb:individual} and
the discussion in the previous section, we 
observe a significant variation amongst
the $a_{0}$ values with lattice spacing, whereas the values of $a_2$  
are approximately constant with lattice spacing. We also recall that the
$a_4$ coefficient was undetermined. This suggests that we should allow
for some variation of the $a_0$ coefficient with lattice spacing and  
in consequence we adopt the following modified version of the 
Adelaide and naive fitting functions,  
based on Eqs.~(\ref{eq:adel}) and (\ref{eq:naive}).
%
%
\be
\sqrt{(M_V^{deg})^{2} - \Sigma_{TOT}} =
(a_0^{cont} + X_1 a + X_2 a^2) +
a_2 (M_{PS}^{deg})^2 + a_4 (M_{PS}^{deg})^4 + a_6 (M_{PS}^{deg})^6,
\label{eq:adel_global} 
\ee
and 
\be
M_V^{deg} =
(a_0^{cont} + X_1 a + X_2 a^2) +
a_2 (M_{PS}^{deg})^2 + a_4 (M_{PS}^{deg})^4 + a_6 (M_{PS}^{deg})^6 \, ,
\label{eq:naive_global}
\ee
%
%
respectively. As in the previous section we refer to the above fits 
as ``cubic'',
since they include the $a_6$ term 
(which is proportional to $m_q^3$). As above we
also perform fits with $a_6$ set to zero, referring to these as
``quadratic''.
These fitting functions, Eqs.~(\ref{eq:adel_global}) and 
(\ref{eq:naive_global}),     
have both ${\cal O}(a)$ and ${\cal O}(a^2)$ terms,  
because the lattice action used is only mean-field improved  
and will contain ${\cal O}(a^2)$ errors, 
together with some residual
errors at ${\cal O}(a)$.
In the following we will experiment by turning off the ${\cal O}(a)$ term
(i.e. by setting $X_1 \equiv 0$) in Eqs.~(\ref{eq:adel_global}) and  
(\ref{eq:naive_global}) in order to see whether 
these residual ${\cal O}(a)$ errors
are significant.
Note that we also included ${\cal O}(a,a^2)$ terms in $a_2$ (and even $a_4$)
as a check but found that these fits were unstable,  
confirming the findings of the previous section that there are
discernible lattice spacing effects only in the $a_0$ coefficient.

The global fits used both $r_0$ and the string tension, $\sigma$, to set
the scale. Thus we have a large number of fit types which are summarised
in Table~\ref{tb:fit_types}. Indeed, there are two
choices from each of the four columns in Table~\ref{tb:fit_types}, 
making a total of $2^4$ fitting procedures.
In the following all $2^4$ fits were performed.
%
%
\begin{table}[*htbp]
\begin{center}
\begin{ruledtabular}
\begin{tabular}{@{}llll@{}}
Approach                     & Chiral Extrapolation               & Treatment of Lattice           & Lattice Spacing \\
                             &                                    & Spacing Artefacts in $a_0$     & set from \\
\hline
Adelaide                     & Cubic                              & $a_0$ term has                  & \multicolumn{1}{c}{$r_0$}    \\
i.e. eq.\ref{eq:adel_global} & i.e. ${\cal O}(M_{PS}^6)$ included & ${\cal O}(a + a^2)$ corrections & \\
\hline
Naive                        & Quadratic                          & $a_0$ term has                  & \multicolumn{1}{c}{$\sigma$} \\
i.e. eq.\ref{eq:naive_global}& i.e. no ${\cal O}(M_{PS}^6)$ term  & only ${\cal O}(a^2)$ corrections& \\
\end{tabular}
\end{ruledtabular}
\end{center}
\caption{The different fit types used in the global analysis.
{}Fits for each of the $2^4$ choices depicted above were performed.
\label{tb:fit_types}}
\end{table}
%
%
As noted above, the global fits contain 80 data points, and the largest
number of fitting parameters is six
($a_0^{cont}, X_1, X_2, a_2, a_4$ and $a_6$)\footnote{The Adelaide approach
also involves the $\Lambda$ parameter which is discussed shortly.}.
Thus the global fits are highly constrained.

\begin{figure}[*htbp]
\begin{center}
\includegraphics[angle=0, width=0.85\textwidth]{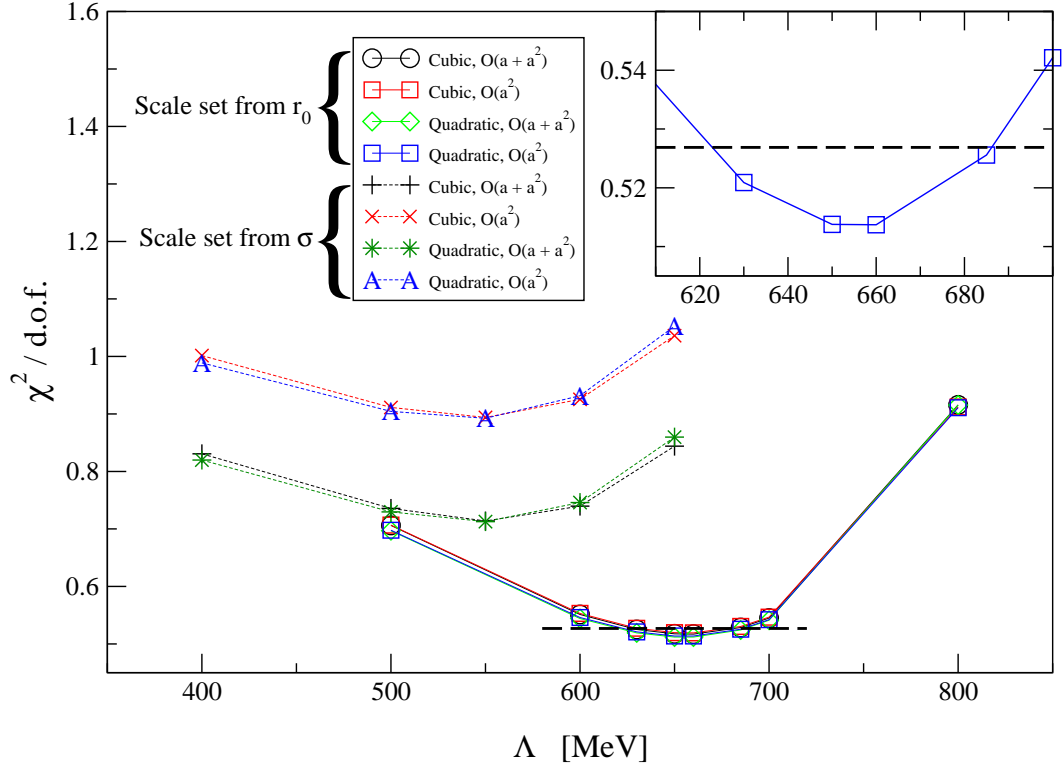}
\caption
{Variation of the $\chi^2 / d.o.f$ against $\Lambda$.
The insert shows a closeup of the minimum for the quadratic fit
with ${\cal O}(a^2)$ errors using $r_0$ to set the scale
(i.e. the preferred fitting procedure).
The dashed horizontal line represents an increase of $\chi^2$ from its
minimum value by unity for this fit procedure (i.e. it represents one
standard deviation) -- see Sec.~\ref{sec:expt}.
The intercept of this dashed line with the $\chi^2$ curves
(at $\Lambda = $620 and 690 MeV) is used to derive the range of acceptable
$\Lambda$ values.
\label{fg:chi}}
\end{center}
\end{figure}

Before presenting results from the global fits, we recall our
discussion of the $\Lambda$ parameter in Sec.~\ref{sec:pq}.
The Adelaide approach motivates the introduction of the mass scale, 
$\Lambda$, as corresponding to the physical size of 
the pion source in the hadron which controls the chiral physics.
It serves to separate the region where chiral physics is important from
that where the internal structure of the hadron, which is not part of
the effective field theory, becomes dominant.
Since we are performing a highly constrained fit procedure,
we are able to derive the best $\Lambda$ value from the data
as follows.
{}Figure~\ref{fg:chi} shows the $\chi^2 / d.o.f.$ as a function
of $\Lambda$ and  
we see that for the results where $r_0$ was used to set
the scale, all four fitting types display near identical $\chi^2$
behaviour, with a distinct minimum at $\Lambda \approx 650$ MeV. In
other words, the $\chi^2$ behaviour is independent of whether we chirally
expand to ${\cal O}(M_{PS}^4)$ or ${\cal O}(M_{PS}^6)$ or whether
we allow for lattice systematics in the $a_0$ coefficient of ${\cal
O}(a+a^2)$ or ${\cal O}(a^2)$.

{}For the case where $\sigma$ is used to set the scale,
Fig.~\ref{fg:chi} shows that there is a distinct minimum at $\Lambda
\approx 550$ MeV for all four
fitting types.
All other things being equal, there
is no preference between the cubic or quadratic chiral fits.  However,
the plot clearly shows that the ${\cal O}(a + a^2)$ fits are favoured,
in that they produce a lower $\chi^2$ value than the pure ${\cal
O}(a^2)$ fits.  As a test, we have also fitted the data with a pure
${\cal O}(a)$ correction (i.e., Eq.~(\ref{eq:adel_global}) with $X_2 =
0$) and found that the $\chi^2$ values for this fit overlay those from
the ${\cal O}(a + a^2)$ fits. This is strongly suggestive that the
dominant lattice-spacing systematic is ${\cal O}(a)$ when the string
tension is used to set the scale.  
%

%

%
%

%

We now use these values of $\Lambda$ (550 MeV and 650 MeV for
the $\sigma$ and $r_0$ cases, respectively) to perform the 16
global fits discussed above and listed in Table~\ref{tb:fit_types}.
Table~\ref{tb:global} gives the coefficients and the $\chi^2/d.o.f.$
for each fit.

\begin{table}[*htbp]
\begin{center}
\begin{ruledtabular}
\begin{tabular}{ccccccccc}
Fit       & Scale & $a_0^{cont}$ &      $X_1$    &      $X_2$    &    $a_2$    &   $a_4$     &    $a_6$    & $\chi^2/d.o.f.$ \\
Approach  & from  & [GeV] & [GeVfm$^{-1}$]& [GeVfm$^{-2}$]& [GeV$^{-1}$]& [GeV$^{-3}$]& [GeV$^{-5}$]&                 \\
\hline
\multicolumn{9}{c}{Cubic chiral extrapolation; $a_0$ contains ${\cal O}(a + a^2)$} \\
Adelaide& $r_0$    & 0.844\err{13}{16} & -0.11\err{15}{13} & -1.1\err{ 3}{ 4} & 0.47\err{ 5}{ 4} & -0.02\err{ 8}{10} & -0.02\err{ 5}{ 4} & 38 / 74 \\
Adelaide& $\sigma$ & 0.836\err{ 9}{11} & -0.37\err{10}{ 9} & -0.2\err{ 2}{ 3} & 0.44\err{ 5}{ 4} &  0.04\err{ 7}{ 9} & -0.06\err{ 5}{ 4} & 53 / 74 \\
Naive   & $r_0$    & 0.819\err{13}{17} & -0.15\err{15}{13} & -1.1\err{ 3}{ 4} & 0.56\err{ 6}{ 5} & -0.16\err{ 8}{10} & 0.05\err{ 5}{ 4} & 77 / 74 \\
Naive   & $\sigma$ & 0.805\err{10}{12} & -0.38\err{11}{ 9} & -0.3\err{ 2}{ 3} & 0.57\err{ 5}{ 4} & -0.18\err{ 8}{10} & 0.06\err{ 6}{ 5} & 73 / 74 \\
\hline
\multicolumn{9}{c}{Cubic chiral extrapolation; $a_0$ contains ${\cal O}(a^2)$ only} \\
Adelaide& $r_0$    & 0.835\err{ 8}{ 9} &---& -1.40\err{ 3}{ 4} & 0.48\err{ 5}{ 4} & -0.03\err{ 8}{10} & -0.02\err{ 5}{ 4} & 39 / 75 \\
Adelaide& $\sigma$ & 0.807\err{ 6}{ 8} &---& -1.24\err{ 3}{ 3} & 0.43\err{ 5}{ 4} &  0.06\err{ 8}{ 9} & -0.06\err{ 5}{ 4} & 67 / 75 \\
Naive   & $r_0$    & 0.806\err{ 8}{10} &---& -1.49\err{ 4}{ 4} & 0.56\err{ 6}{ 5} & -0.17\err{ 8}{10} &  0.06\err{ 5}{ 4} & 78 / 75 \\
Naive   & $\sigma$ & 0.775\err{ 7}{ 8} &---& -1.31\err{ 4}{ 4} & 0.56\err{ 5}{ 4} & -0.16\err{ 8}{10} &  0.05\err{ 5}{ 5} & 87 / 75 \\
\hline
\multicolumn{9}{c}{Quadratic chiral extrapolation; $a_0$ contains ${\cal O}(a + a^2)$} \\
Adelaide& $r_0$    & 0.840\err{10}{12} & -0.11\err{14}{13} & -1.1\err{ 3}{ 4} & 0.493\err{12}{11} & -0.061\err{ 8}{ 9} &---& 38 / 75 \\
Adelaide& $\sigma$ & 0.829\err{ 8}{ 9} & -0.37\err{10}{ 9} & -0.2\err{ 2}{ 3} & 0.490\err{13}{11} & -0.052\err{10}{11} &---& 54 / 75 \\
Naive   & $r_0$    & 0.828\err{11}{13} & -0.16\err{15}{13} & -1.1\err{ 3}{ 4} & 0.505\err{13}{11} & -0.068\err{ 9}{10} &---& 78 / 75 \\
Naive   & $\sigma$ & 0.812\err{ 8}{ 9} & -0.37\err{11}{ 9} & -0.3\err{ 2}{ 3} & 0.523\err{13}{12} & -0.075\err{11}{11} &---& 74 / 75 \\
\hline
\multicolumn{9}{c}{Quadratic chiral extrapolation; $a_0$ contains ${\cal O}(a^2)$ only} \\
Adelaide& $r_0$    & 0.832\err{ 4}{ 4} &---& -1.40\err{ 3}{ 4} & 0.494\err{12}{11} & -0.061\err{ 8}{ 9} &---& 39 / 76 \\
Adelaide& $\sigma$ & 0.799\err{ 3}{ 4} &---& -1.23\err{ 3}{ 3} & 0.486\err{13}{11} & -0.046\err{10}{11} &---& 68 / 76 \\
Naive   & $r_0$    & 0.815\err{ 4}{ 4} &---& -1.49\err{ 4}{ 4} & 0.506\err{12}{11} & -0.068\err{ 8}{10} &---& 79 / 76 \\
Naive   & $\sigma$ & 0.781\err{ 3}{ 4} &---& -1.31\err{ 3}{ 4} & 0.520\err{13}{12} & -0.069\err{11}{11} &---& 88 / 76 \\
\hline
\end{tabular}
\end{ruledtabular}
\end{center}
\caption{The results of the global fit analysis. Fits for all $2^4$
combinations depicted in Table~\ref{tb:fit_types} are shown.
\label{tb:global}}
\end{table}
Summarising the results of Table~\ref{tb:global} and Fig.~\ref{fg:chi}
(and referring to the $2^4$ different fit types listed in Table
\ref{tb:fit_types}) we note:

{\em Fit Approach:} The Adelaide method always gives a smaller
$\chi^2$ than the naive approach, confirming it as the preferred chiral
extrapolation procedure.

{\em Chiral Extrapolation:} The cubic chiral extrapolation
(i.e., introducing the ${\cal O}(M_{PS}^6)$ term in
Eqs.~(\ref{eq:adel_global}) and  (\ref{eq:naive_global}) ) 
leads to a poorly determined $a_6$ coefficient in all cases. Furthermore
it causes the $a_4$ coefficient to become much more poorly determined
than it is in the quadratic chiral extrapolation cases.

{\em Treatment of Lattice Spacing Artefacts in $a_0$:}
{}From Table~\ref{tb:global}, when the scale is set from
$r_0$, the $a_0$ and $a_2$ coefficients
(which are the dominant terms in the chiral extrapolation) do not
depend on whether ${\cal O}(a+a^2)$ or ${\cal O}(a^2)$--only
corrections are applied to the $a_0$ coefficient.
Indeed, when ${\cal O}(a^2)$ only corrections are used,
the error in $a_0$ is reduced.
However, when the scale is set from $\sigma$, $a_0$ {\em does}
depend on how the ${\cal O}(a)$ systematics are treated.
Note also that $X_1$ (i.e. the ${\cal O}(a)$ coefficient) in the $\sigma$
fits are 2--3 times larger than those from the $r_0$ fits.
This supports our earlier comments above regarding the probable
${\cal O}(a)$ systematics when the $\sigma$ scale was used.

{\em Quantity used to set Lattice Spacing:}
In the Adelaide approach, setting the scale from $r_0$ gives
a significantly smaller $\chi^2$ than using $\sigma$
(see Fig.~\ref{fg:chi}).
Given this, and the comments above regarding the probable
${\cal O}(a)$ systematics in the $\sigma$ data,
we use $r_0$ as our preferred method for setting the scale.
In the naive case the data does not favour
setting the scale from either $r_0$ or $\sigma$.

On the basis of these results, we choose the quadratic chiral extrapolation
method with the scale set from $r_0$ and ${\cal O}(a^2)$ corrections in
the $a_0$ coefficient to define the central value of both the Adelaide
and naive fitting procedure.
The spread from the other fitting types is used to define the error.
In Sec.~\ref{sec:expt}, we determine the predictions for physical
$\rho-$meson mass from these fitting types.




\section{Physical Predictions}
\label{sec:expt}

We are now ready to estimate $M_\rho$ in the
continuum using the Adelaide and naive fits performed in the previous section.
We obtain this prediction from
Eqs.~(\ref{eq:adel})
and (\ref{eq:naive}) by setting
$M_{PS}^{deg} = M_{PS}^{non-deg} = M_{PS}^{unit} = \mu_\pi$.
We set $\Delta M_{\omega\rho}$ to zero in this calculation and also 
note that $\Sigma^\rho_{\eta' \rho}$ vanishes, as required.
Since we are predicting the continuum value for the vector meson mass,
we calculate the integrals in Eqs.~(\ref{eq:sigma_pipi}) and  
(\ref{eq:sigma_piomega})
directly, rather than using the lattice interpretation of the integral
in Eq.~(\ref{eq:lat_int}).

We obtain continuum estimates of all $2^4$ fitting types (see Table
\ref{tb:fit_types}) using the coefficients, $a_0^{cont}$ and $a_{2,4,6}$, of
the global fits of Sec.~\ref{sec:global} (i.e., 
those in Table~\ref{tb:global}).
Table~\ref{tb:mass_estimates} displays these mass predictions.
{}For the Adelaide case we use
$\Lambda = 650(550)$ MeV, with the physical scale set using
$r_0 (\sigma)$ (see Sec.~\ref{sec:global}).
Note that we use the global analysis 
(Sec.~\ref{sec:global}), rather than the
analysis of Sec.~\ref{sec:individual}, which treated the ensembles
separately, since the global fit is much more tightly constrained.

\begin{figure}[*htbp]
\begin{center}
\includegraphics[angle=0, width=0.85\textwidth]{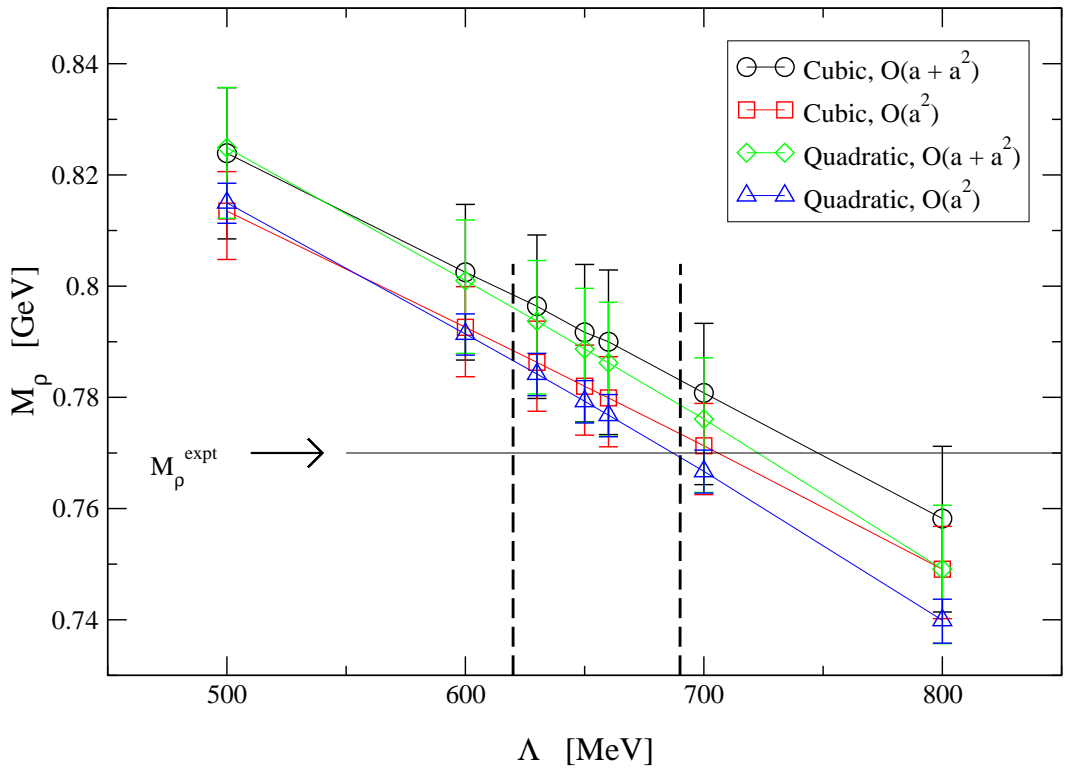}
\caption
{Variation of the physical mass of the $\rho$-meson, $M_\rho$, 
as a function of $\Lambda$, using the Adelaide approach
with the scale set from $r_0$.
The two vertical dashed lines define the range of acceptable $\Lambda$
values (620 MeV $\le \Lambda \le$ 690 MeV) obtained by increasing
$\chi^2$ by unity in Fig.~\ref{fg:chi}.
\label{fg:mrho_v_lambda}}
\end{center}
\end{figure}

It is interesting to study the variation in the 
prediction of the physical mass of the $\rho$ 
with $\Lambda$.
{}Figure~\ref{fg:mrho_v_lambda} shows how the $M_\rho$ prediction varies
with $\Lambda$ for each of the Adelaide fits based on the $r_0$ scale.
Using the $\chi^2$ plot in Fig.~\ref{fg:chi}, we can estimate the
range of acceptable $\Lambda$ values defined by increasing $\chi^2$
by unity from its minimum, which represents one standard deviation.
The horizontal dashed line in Fig.~\ref{fg:chi} lies along $\chi^2$
values one more than the minimum for the $r_0$ case.
{}From this, we determine that the acceptable range of $\Lambda$ values
is $620\le \Lambda \le 690\mev$ (see the insert graph
in Fig.~\ref{fg:chi}).
This range of $\Lambda$ values is depicted by the vertical dashed lines
in Fig.~\ref{fg:mrho_v_lambda}.


\begin{table}[*htbp]
\begin{center}
\begin{ruledtabular}
\begin{tabular}{clll}
Source & Fit      & Scale    & \multicolumn{1}{c}{M$_{\rho}$} \\
       & Procedure& from     & \multicolumn{1}{c}{[GeV]}      \\
\hline
Experiment& &         &       0.770       \\
\hline
&\multicolumn{3}{c}{Cubic chiral extrapolation; $a_0$ contains ${\cal O}(a + a^2)$} \\
Dynamical& Adelaide & $r_0$    & 0.792\err{12}{16} \\
''       & Adelaide & $\sigma$ & 0.810\err{ 9}{11} \\
''       & Naive    & $r_0$    & 0.829\err{12}{16} \\
''       & Naive    & $\sigma$ & 0.815\err{ 9}{12} \\
\hline
&\multicolumn{3}{c}{Cubic chiral extrapolation; $a_0$ contains ${\cal O}(a^2)$ only} \\
Dynamical& Adelaide & $r_0$    & 0.782\err{ 7}{ 9} \\
''       & Adelaide & $\sigma$ & 0.781\err{ 6}{ 7} \\
''       & Naive    & $r_0$    & 0.817\err{ 7}{ 9} \\
''       & Naive    & $\sigma$ & 0.786\err{ 6}{ 7} \\
\hline
&\multicolumn{3}{c}{Quadratic chiral extrapolation; $a_0$ contains ${\cal O}(a + a^2)$} \\
Dynamical& Adelaide & $r_0$    & 0.789\err{11}{13} \\
''       & Adelaide & $\sigma$ & 0.805\err{ 8}{ 9} \\
''       & Naive    & $r_0$    & 0.837\err{11}{13} \\
''       & Naive    & $\sigma$ & 0.822\err{ 8}{ 9} \\
\hline
&\multicolumn{3}{c}{Quadratic chiral extrapolation; $a_0$ contains ${\cal O}(a^2)$ only}\\
Dynamical& Adelaide & $r_0$    & 0.779\err{ 4}{ 4} \\
''       & Adelaide & $\sigma$ & 0.774\err{ 3}{ 3} \\
''       & Naive    & $r_0$    & 0.825\err{ 4}{ 4} \\
''       & Naive    & $\sigma$ & 0.791\err{ 3}{ 3} \\
\end{tabular}
\end{ruledtabular}
\end{center}
\caption{Estimates of M$_{\rho}$ obtained from the global fits.
The Adelaide fits used $\Lambda = $ 650(550)~MeV when the scale
is set from $r_0(\sigma)$.
\label{tb:mass_estimates}}
\end{table}
%

%

%
%

%

{}From a detailed examination of Table~\ref{tb:mass_estimates}
and Fig.~\ref{fg:mrho_v_lambda} we draw the following conclusions:
\begin{itemize}

\item The (statistical) errors in the mass estimates are typically
around 1\%.

\item The Adelaide fitting procedure is very stable when the scale
is set using $r_0$. When the scale is taken from the string tension,
the four Adelaide fits are not in mutual agreement. The probable reason
for this, as outlined in Sec.~\ref{sec:global}, is that using $\sigma$
to set the scale introduces significant ${\cal O}(a)$ errors.

\item The Adelaide fitting procedure is quite accurate -- lying at most
at twice the {\em statistical} standard error from the experimental
value for the $r_0$ case.
It would require an uncertainty of only around 1-2\% in $r_0$ (and 
around 2-6\% in $\sqrt{\sigma}$) for the Adelaide central values to be in
agreement with the experimental value of $M_\rho$.

\item From Fig.~\ref{fg:mrho_v_lambda}, the variation of $M_\rho$
with $\Lambda$ is small - roughly the same order as the other
uncertainties.

\item The naive fitting procedure has both a larger spread of values
and is further from the experimental value than the Adelaide procedure.
\end{itemize}

As a result, we conclude that the Adelaide procedure represents a significant
improvement over the naive approach.

We obtain final estimates of $M_\rho$ by taking the the quadratic
chiral extrapolation, the scale set from $r_0$, and ${\cal O}(a^2)$
corrections in the $a_0$ coefficient in both the Adelaide and naive
fitting procedure. (The reason for this choice of fit type is
described in detail in Sec.~\ref{sec:global}.)
The central value in the Adelaide case was obtained at $\Lambda=$ 655~MeV
from Fig.~\ref{fg:mrho_v_lambda} (which is an adjustment of 1 MeV
from the value obtained at $\Lambda=$ 650~MeV in Table~\ref{tb:mass_estimates}).
The value of $\Lambda =$ 655~MeV was used since it is where the minimum
of $\chi^2$ occurs in Fig.~\ref{fg:chi}.
We obtain an estimate of the error in the fit procedure from
the spread in the mass predictions using $r_0$ for the scale
(since we have reservations about the method when the string tension
is used to set the scale).
An estimate of the uncertainty associated with the determination of $\Lambda$ 
comes from varying $\chi^2$ by unity as described above --- i.e.
by reading off this error from the vertical dashed lines
in Fig.~\ref{fg:mrho_v_lambda}.
{}Finally then we are led to the following result for the physical mass
of the $\rho$-meson:
\bea
\label{eq:mass_final_adel}
M_\rho^{Adelaide} &=& 778(4)\er{16}{6}\er{8}{9} \textrm{MeV,} \\
\label{eq:mass_final_naive}
M_\rho^{Naive}    &=& 825(4)\er{12}{8} \textrm{MeV,}
\eea
where the first error is statistical, the second is from the fit
procedure, and, in the Adelaide case, the third error is from the
determination of $\Lambda$. The second error on the Adelaide result
also includes an uncertainty from the choice of finite-range
regulator, which contributes $+3/-6\mev$ to the error
\cite{Allton:2005fb}.  Note that we have not included an error from
the determination of $r_0$ itself.
The only other effect which separates our analysis from nature
is that the data we are analysing contains only 2 rather than 2+1
light dynamical flavours. We have no way to estimate the residual
systematic error from this source once $r_0$ is matched to the physical
value.

The fit parameters shown in Table~\ref{tb:global} allow one to shift
each of the simulation results to the infinite-volume, continuum limit
and to remove the effects of partial quenching --- hence restoring
unitarity in the quark masses. This can be achieved by first
considering the preferred form of Eq.~(\ref{eq:adel_global}), where
$X_1=a_6=0$,
\begin{equation}
\sqrt{(M_V^{deg})^{2} - \Sigma_{TOT}} =
(a_0^{cont} + X_2 a^2) +
a_2 (M_{PS}^{deg})^2 + a_4 (M_{PS}^{deg})^4\, .
\label{eq:global_pref}
\end{equation}
This can then be used to let $M_V^{deg}\to M_V^{unit}$ at the same
time as removing volume and discretisation artifacts.
Rewriting Eq.~(\ref{eq:global_pref}) in terms of the fit vector meson mass
gives,
\begin{equation}
M_V^{deg}(a) = \bigg[\left((a_0^{cont} + X_2 a^2) +
a_2 (M_{PS}^{deg})^2 + a_4 (M_{PS}^{deg})^4 \right)^2 + \Sigma_{TOT}(M_{PS}^{deg};L)\bigg]^{1/2} \, .
\end{equation}
In the physical continuum limit, this becomes
\begin{eqnarray}
M_V^{unit}(a\to 0) = \bigg[\left(a_0^{cont} +
a_2 (M_{PS}^{unit})^2 + a_4 (M_{PS}^{unit})^4 \right)^2 + \Sigma_{TOT}(M_{PS}^{unit};L\to\infty)\bigg]^{1/2} \, .
\end{eqnarray}
With the unknown parameters determined from the best fit to the entire
data set, this then provides a prescription for restoring the physical
limit of the data. Specifically, each of the vector meson masses
(calculated at finite $a$ and $L$) are shifted by an amount
\begin{eqnarray}
\delta M &=& \bigg[\left(a_0^{cont} +
a_2 (M_{PS}^{unit})^2 + a_4 (M_{PS}^{unit})^4 \right)^2
 + \Sigma_{TOT}(M_{PS}^{unit};L\to\infty)\bigg]^{1/2} \non\\
& &-\bigg[\left((a_0^{cont} + X_2 a^2) +
a_2 (M_{PS}^{deg})^2 + a_4 (M_{PS}^{deg})^4 \right)^2
 + \Sigma_{TOT}(M_{PS}^{deg};L)\bigg]^{1/2} \, .
\end{eqnarray}
Here it is understood that the pion masses that one should therefore
plot against on the $x$-axis are the unitary pion, at the point where
the {\it sea} mass is held fixed and the {\it valence} mass is
changed, ie.~$M(\beta,\kappa_{\rm sea};\kappa_{\rm val},\kappa_{\rm
val})\to M(\beta,\kappa_{\rm sea};\kappa_{\rm sea},\kappa_{\rm sea})$.
The final estimate of the physical point is given by
\begin{equation}
\left[M_V^{unit}(a\to 0;L\to\infty)\right]_{\rm estimate} = \left[M_V^{deg}\right]_{\rm lattice} + \delta M \, .
\end{equation}

\begin{figure}[*htbp]
\begin{center}
\includegraphics[angle=0, width=0.85\textwidth]{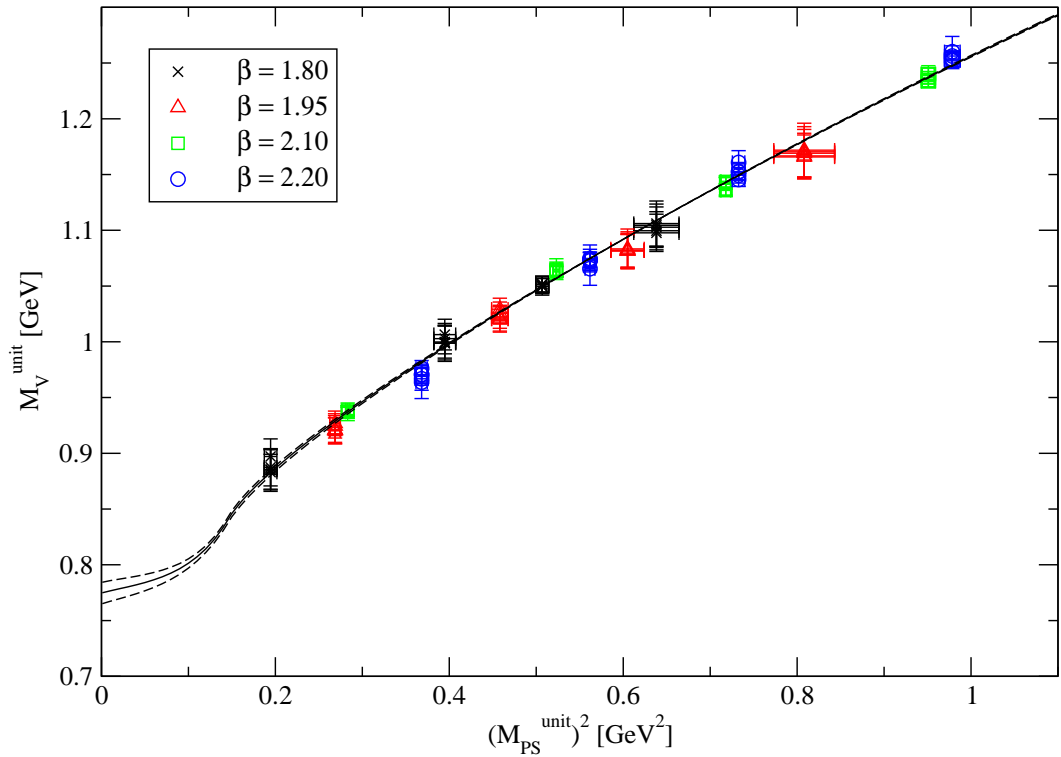}
\caption{
The same 80 lattice data points as in Fig.~\protect\ref{fg:global},
after correction to restore the infinite-volume, continuum and
quark-mass unitarity limits.  The central curve displays the best-fit
from the global analysis with $\Lambda =$ 655 MeV.  
The dashed curves show the (statistically constrained) bounds
on the FRR scale, $0.620<\Lambda<0.690$ GeV.
\label{fg:famous}}
\end{center}
\end{figure}

The results of these shifts are displayed in Fig.~\ref{fg:famous},
where we observe a remarkable result.  The tremendous spread of data
seen in Fig.~\ref{fg:global} is dramatically reduced, with all 80
points now lying very accurately on a universal curve.


%
%

%

One feature of Fig.~\ref{fg:global} is the very steady approach to the
chiral limit from the physical point. This yields a value for the
vector meson mass in the chiral limit ($M_\rho^0$) that is very near
the physical value. To remove the overall scale, we report on the mass
shift between the physical and chiral values, finding
\begin{equation}
M_\rho - M_\rho^0 = 3.7(2)(4)(8) \mev ,
\end{equation}
where the sources of uncertainty are the same as those reported above
for $M_\rho$ in Eq.~(\ref{eq:mass_final_adel}). This small number
indicates that the sigma term for the $\rho$ is significantly smaller
than the nucleon, where the derivative is observed to increase near
the chiral limit \cite{Leinweber:2003dg}. The reduced slope in the
case of the $\rho$ arises from the presence of substantial spectral
strength in the low-energy two-pion channel below the $\rho$-meson
mass \cite{Leinweber:1993yw}.



\section{Conclusions}

In summary, we have tackled the ambitious task of producing a single,
unified chiral fit to all of the accurate CP-PACS data for the mass of
the $\rho$-meson in partially quenched QCD -- {\it i.e.}, the case
where $\ksea \ne \kval$.  As well as using a naive polynomial fit in
$M_{PS}^2$, we have generalized the Adelaide approach, developing a
unified analysis approach, to fit the data.  This approach enables one
to account for finite volume errors by evaluating the chiral
self-energy contributions with the same momentum discretisation
implicit in the lattice simulations.  In addition, we have been able
to quantify the residual ${\cal O}(a)$ effects and hence carry out a
continuum extrapolation (c.f. Figs.~\ref{fg:a0_cont} and
\ref{fg:a2_cont}).

The $\chi^2$ obtained for the Adelaide fit, with the physical scale
set using $r_0$, is a factor of two lower than that for any other
method.  This provides considerable confidence in the method, even
before it is used to produce the physical mass of the $\rho$. The
quality of the fit leads to very small (statistical) error bars ---
see Table~\ref{tb:mass_estimates}.  

In addition, it is possible to estimate the systematic errors in the
extrapolation to the physical $\rho-$mass associated with the fitting
procedure (both the chiral and continuum-limit fitting procedures).
In particular, the finite-range regulator parameter, $\Lambda$, is
constrained by the model-independent lattice QCD data, and the
variation of the $M_\rho$ prediction within this range was found to be
1\%.  

The curve through Fig.~\ref{fg:famous} displays the determined
variation of the $\rho$-meson mass with pion mass.  This curve also
presents an extrapolation to the physical point, allowing extraction
of the physical $\rho$-meson mass
\begin{equation}
M_\rho = 778(4)\er{16}{6}\er{8}{9} \mev, \\
\end{equation}
where the first error is statistical, the second is from variations of
the fit procedure and the third from the determination of
$\Lambda$. 
Whereas the naive fitting procedure leads to a value that is 50-60 MeV
too high, the result from the chiral analysis is in excellent
agreement with the experimentally observed mass.


\section*{Acknowledgements}
CRA and WA would like to thank the CSSM for their support and kind
hospitality.  WA would like to thank PPARC for travel support.  The
authors would like to thank Stewart Wright and Graham Shore for
helpful comments.  This work was supported by the Australian Research
Council and by DOE contract DE-AC05-84ER40150, under which SURA
operates Jefferson Laboratory.





\begin{thebibliography}{99}

\bibitem{cppacs}
A.~Ali Khan {\it et al.}  [CP-PACS Collaboration],
Phys.\ Rev.\ D {\bf 65}, 054505 (2002)
[Erratum-ibid.\ D {\bf 67}, 059901 (2003)]
[arXiv:hep-lat/0105015].


\bibitem{Leinweber:2003dg}
D.~B.~Leinweber, A.~W.~Thomas and R.~D.~Young,
Phys.\ Rev.\ Lett.\  {\bf 92} (2004) 242002
[arXiv:hep-lat/0302020].

\bibitem{Procura:2003ig}
M.~Procura, T.~R.~Hemmert and W.~Weise,
Phys.\ Rev.\ D {\bf 69}, 034505 (2004)
[arXiv:hep-lat/0309020].

\bibitem{Leinweber:1999ig}
D.~B.~Leinweber, A.~W.~Thomas, K.~Tsushima and S.~V.~Wright,
Phys.\ Rev.\ D {\bf 61}, 074502 (2000)
[arXiv:hep-lat/9906027].

\bibitem{Durr:2002zx}
  S.~Durr,
  Eur.\ Phys.\ J.\ C {\bf 29}, 383 (2003)
  [arXiv:hep-lat/0208051].

\bibitem{Bernard:2002yk}
C.~Bernard, S.~Hashimoto, D.~B.~Leinweber, P.~Lepage, E.~Pallante, S.~R.~Sharpe and H.~Wittig,
Nucl.\ Phys.\ Proc.\ Suppl.\  {\bf 119}, 170 (2003)
[arXiv:hep-lat/0209086].

\bibitem{Young:2002ib}
R.~D.~Young, D.~B.~Leinweber and A.~W.~Thomas,
Prog.\ Part.\ Nucl.\ Phys.\  {\bf 50}, 399 (2003)
[arXiv:hep-lat/0212031].

\bibitem{Beane:2004ks}
  S.~R.~Beane,
  Nucl.\ Phys.\ B {\bf 695}, 192 (2004)
  [arXiv:hep-lat/0403030].

\bibitem{Thomas:2004iw}
  A.~W.~Thomas, P.~A.~M.~Guichon, D.~B.~Leinweber and R.~D.~Young,
  Prog.\ Theor.\ Phys.\ Suppl.\  {\bf 156}, 124 (2004)
  [arXiv:nucl-th/0411014].

\bibitem{Leinweber:2005xz}
  D.~B.~Leinweber, A.~W.~Thomas and R.~D.~Young,
  Nucl.\ Phys.\ A {\bf 755}, 59 (2005)
  [arXiv:hep-lat/0501028].

\bibitem{Donoghue:1998bs}
J.~F.~Donoghue, B.~R.~Holstein and B.~Borasoy,
Phys.\ Rev.\ D {\bf 59}, 036002 (1999)
[arXiv:hep-ph/9804281].

\bibitem{Djukanovic:2005jy}
  D.~Djukanovic, M.~R.~Schindler, J.~Gegelia and S.~Scherer,
  Phys.\ Rev.\ D {\bf 72}, 045002 (2005).

\bibitem{Young:2004tb}
  R.~D.~Young, D.~B.~Leinweber and A.~W.~Thomas,
  Phys.\ Rev.\ D {\bf 71}, 014001 (2005)
  [arXiv:hep-lat/0406001].

\bibitem{Golterman:1997st}
  M.~F.~L.~Golterman and K.~C.~L.~Leung,
  Phys.\ Rev.\ D {\bf 57}, 5703 (1998)
  [arXiv:hep-lat/9711033].

\bibitem{Sharpe:2001fh}
S.~R.~Sharpe and N.~Shoresh,
Phys.\ Rev.\ D {\bf 64}, 114510 (2001)
[arXiv:hep-lat/0108003].

\bibitem{Chen:2001yi}
  J.~W.~Chen and M.~J.~Savage,
  Phys.\ Rev.\ D {\bf 65}, 094001 (2002)
  [arXiv:hep-lat/0111050].

\bibitem{Beane:2002vq}
  S.~R.~Beane and M.~J.~Savage,
  Nucl.\ Phys.\ A {\bf 709}, 319 (2002)
  [arXiv:hep-lat/0203003].

\bibitem{Leinweber:2002qb}
  D.~B.~Leinweber,
  Phys.\ Rev.\ D {\bf 69}, 014005 (2004)
  [arXiv:hep-lat/0211017].

\bibitem{Arndt:2003ww}
  D.~Arndt and B.~C.~Tiburzi,
  Phys.\ Rev.\ D {\bf 68}, 094501 (2003)
  [arXiv:hep-lat/0307003].

\bibitem{Arndt:2004bg}
  D.~Arndt and C.-J.~D.~Lin,
  Phys.\ Rev.\ D {\bf 70}, 014503 (2004)
  [arXiv:hep-lat/0403012].

\bibitem{Bijnens:2004hk}
  J.~Bijnens, N.~Danielsson and T.~A.~Lahde,
  Phys.\ Rev.\ D {\bf 70}, 111503 (2004)
  [arXiv:hep-lat/0406017].

\bibitem{Detmold:2005pt}
  W.~Detmold and C.-J.~D.~Lin,
  Phys.\ Rev.\ D {\bf 71}, 054510 (2005)
  [arXiv:hep-lat/0501007].

\bibitem{Bijnens:1997ni}
  J.~Bijnens, P.~Gosdzinsky and P.~Talavera,
  Nucl.\ Phys.\ B {\bf 501}, 495 (1997)
  [arXiv:hep-ph/9704212].

\bibitem{Bruns:2004tj}
  P.~C.~Bruns and U.~G.~Meissner,
  Eur.\ Phys.\ J.\ C {\bf 40}, 97 (2005)
  [arXiv:hep-ph/0411223].

\bibitem{Allton:2005fb}
  C.~R.~Allton, W.~Armour, D.~B.~Leinweber, A.~W.~Thomas and R.~D.~Young,
  Phys.\ Lett.\ B {\bf 628}, 125 (2005)
  [arXiv:hep-lat/0504022].

\bibitem{Leinweber:1993yw}
D.~B.~Leinweber and T.~D.~Cohen,
Phys.\ Rev.\ D {\bf 49} (1994) 3512
[arXiv:hep-ph/9307261].

\bibitem{adel_rho}
D.~B.~Leinweber, A.~W.~Thomas, K.~Tsushima and S.~V.~Wright,
Phys.\ Rev.\ D {\bf 64}, 094502 (2001)
[arXiv:hep-lat/0104013].

\bibitem{Labrenz:1996jy}
J.~N.~Labrenz and S.~R.~Sharpe,
Phys.\ Rev.\ D {\bf 54}, 4595 (1996)
[arXiv:hep-lat/9605034].

\bibitem{Chow:1997dw}
  C.~K.~Chow and S.~J.~Rey,
  Nucl.\ Phys.\ B {\bf 528}, 303 (1998)
  [arXiv:hep-ph/9708432].

\bibitem{Young:2002cj}
R.~D.~Young, D.~B.~Leinweber, A.~W.~Thomas and S.~V.~Wright,
Phys.\ Rev.\ D {\bf 66}, 094507 (2002)
[arXiv:hep-lat/0205017].

\bibitem{sommer}
R.~Sommer,
Nucl.\ Phys.\ B {\bf 411}, 839 (1994)
[arXiv:hep-lat/9310022].
R.~G.~Edwards, U.~M.~Heller and T.~R.~Klassen,
Nucl.\ Phys.\ B {\bf 517}, 377 (1998)
[arXiv:hep-lat/9711003].

\bibitem{scaling}
R.~G.~Edwards, U.~M.~Heller and T.~R.~Klassen,
Phys.\ Rev.\ Lett.\  {\bf 80}, 3448 (1998)
[arXiv:hep-lat/9711052].
J.~M.~Zanotti, B.~Lasscock, D.~B.~Leinweber and A.~G.~Williams,
Phys.\ Rev.\ D {\bf 71}, 034510 (2005)
[arXiv:hep-lat/0405015].

\bibitem{Bar:2004xp}
  O.~B\"ar,
  Nucl.\ Phys.\ Proc.\ Suppl.\  {\bf 140}, 106 (2005)
  [arXiv:hep-lat/0409123].

\bibitem{Rupak:2002sm}
  G.~Rupak and N.~Shoresh,
  Phys.\ Rev.\ D {\bf 66}, 054503 (2002)
  [arXiv:hep-lat/0201019].

\bibitem{Bar:2003mh}
  O.~Bar, G.~Rupak and N.~Shoresh,
  Phys.\ Rev.\ D {\bf 70}, 034508 (2004)
  [arXiv:hep-lat/0306021].

\bibitem{Aoki:2003yv}
  S.~Aoki,
  Phys.\ Rev.\ D {\bf 68}, 054508 (2003)
  [arXiv:hep-lat/0306027].

\bibitem{Beane:2003xv}
  S.~R.~Beane and M.~J.~Savage,
  Phys.\ Rev.\ D {\bf 68}, 114502 (2003)
  [arXiv:hep-lat/0306036].

\bibitem{Grigoryan:2005zj}
  H.~R.~Grigoryan and A.~W.~Thomas,
  arXiv:hep-lat/0507028.

\end{thebibliography}
\end{document}